%% file: fp0047.tex
\documentclass[sigconf, anonymous=false]{acmart}

\usepackage{balance}
\usepackage{array}
\newcolumntype{L}{>{\centering\arraybackslash}p{0.140\linewidth}}

\usepackage{balance}
\def\BibTeX{{\rm B\kern-.05em{\sc i\kern-.025em b}\kern-.08emT\kern-.1667em\lower.7ex\hbox{E}\kern-.125emX}}

\usepackage{multirow}
\usepackage{calc}
\usepackage{siunitx}
\usepackage[super]{nth}
\usepackage[flushleft]{threeparttable}
\usepackage{enumerate}
\usepackage[shortlabels]{enumitem}

\usepackage{graphics}
\usepackage{cases}
\usepackage{booktabs} % For formal tables
\usepackage{tikz}
\usepackage{natbib}
\usetikzlibrary{bayesnet}
\usepackage{algorithm}
\usepackage{multirow}
\usepackage{amsmath}
\usepackage[noend]{algpseudocode}
\usepackage{subfigure}
\usepackage{cleveref}
\usepackage{dblfloatfix}
\usepackage{bbm}
\usepackage{enumitem}
\usepackage{xfrac}
\usepackage{nth}

\usepackage{mathtools}

\DeclarePairedDelimiterX{\infdivx}[2]{(}{)}{%
  #1\;\delimsize\|\;#2%
}
\newcommand{\infdiv}{D\infdivx}

\usepackage{url}
\setcopyright{acmcopyright}
\copyrightyear{2020} 
\acmYear{2020} 
\setcopyright{acmcopyright}\acmConference[SIGIR '20]{Proceedings of the 43rd International ACM SIGIR Conference on Research and Development in Information Retrieval}{July 25--30, 2020}{Virtual Event, China}
\acmBooktitle{Proceedings of the 43rd International ACM SIGIR Conference on Research and Development in Information Retrieval (SIGIR '20), July 25--30, 2020, Virtual Event, China}
\acmPrice{15.00}
\acmDOI{10.1145/3397271.3401078}
\acmISBN{978-1-4503-8016-4/20/07}

\begin{document}

\fancyhead{}

\title{Transfer Learning via Contextual Invariants for One-to-Many Cross-Domain Recommendation}
%\titlenote{Produces the permission block, and
% copyright information}
%\subtitle{Extended Abstract}
%\subtitlenote{The full version of the author's guide is available as
% \texttt{acmart.pdf} document}

% Removes ACM REF AND copyright block
%\settopmatter{printacmref=false} % Removes citation information below abstract
%\renewcommand\footnotetextcopyrightpermission[1]{} % removes footnote with conference information in first column
\pagestyle{plain} % removes running headers

%\author{Author 2}
%%\authornote{The secretary disavows any knowledge of this author's actions.}
%\affiliation{%
%  \institution{Institute for Clarity in Documentation}
%  \streetaddress{P.O. Box 1212}
%  \postcode{4

%\author{Author 2}
%%\authornote{The secretary disavows any knowledge of this author's actions.}
%\affiliation{%
%  \institution{Institute for Clarity in Documentation}
%  \streetaddress{P.O. Box 1212}
%  \postcode{4

\author{Adit Krishnan$^\dagger$, Mahashweta Das$^*$, Mangesh Bendre$^*$, Hao Yang$^*$, Hari Sundaram$^\dagger$}
\affiliation{
  \institution{$^\dagger$University of Illinois at Urbana-Champaign, IL, USA \\ $^*$ Visa Research, Palo Alto CA, USA}
}
\email{{aditk2, hs1}@illinois.edu}
\email{{mahdas, mbendre, haoyang} @visa.com}
\renewcommand{\authors}{Adit Krishnan, Mahashweta Das, Mangesh Bendre, Hao Yang, Hari Sundaram}
\pagestyle{plain}

\begin{abstract}
\renewcommand*{\thefootnote}{\arabic{footnote}}
The rapid proliferation of new users and items on the social web has aggravated the gray-sheep user/long-tail item challenge in recommender systems. Historically, cross-domain co-clustering methods have successfully leveraged shared users and items across dense and sparse domains to improve inference quality. However, they rely on shared rating data and cannot scale to multiple sparse target domains (i.e., the one-to-many transfer setting). This, combined with the increasing adoption of neural recommender architectures, motivates us to develop scalable neural layer-transfer approaches for cross-domain learning. Our key intuition is to guide neural collaborative filtering with domain-invariant components shared across the dense and sparse domains, improving the user and item representations learned in the sparse domains. We leverage contextual invariances across domains to develop these shared modules, and demonstrate that with user-item interaction context, we can \textit{learn-to-learn} informative representation spaces even with sparse interaction data. We show the effectiveness and scalability of our approach on two public datasets and a massive transaction dataset from Visa, a global payments technology company (19\% Item Recall, 3x faster vs. training separate models for each domain). Our approach is applicable to both \textit{implicit} and \textit{explicit} feedback settings.

\end{abstract}

\maketitle
%\vspace{-9pt}
\input{introduction}

\input{related}
%% \newpage
\input{problemdef}

\input{model}

\input{model_adaptation}
\input{experiments}
\section{Conclusion}
\label{sec:Conclusion}
%This paper proposes an approach to integrate the social network between users to improve the quality of their recommendations. Unlike prior work, we adopt a modular architecture-agnostic framework enabling a broad range of recommender applications. Further, we show that a direct application of metric learning approaches or equivalent formulations could result in interest space collapse. Our pair weighting approach enhances the expressivity of the user interest space and permits for contextual integration of their social structure. Extensive experimental results over five real-world datasets reveal the strengths of our approach.
%
%We identify three rewarding future directions - developing smart samplers to produce informative \emph{fake-pairs} to regularize the interest space, enhancing contextual weighting with a fine-grained combination of the context projections, and finally, developing efficient and expressive discriminator architectures.

This paper proposes a novel approach to address the sparsity problem in the cross-domain setting. We leverage the strengths of meta-transfer learning, grounded on an expressive context pooling strategy to learn effective invariants. Our approach is highly scalable in the one-to-many setting, incurs minimal costs to accommodate new sparse target domains, and significantly outperforms competing methods in our experimental results. A few interesting future directions include updating representation with streaming data and incorporating knowledge priors on expected behavior patterns (e.g., if we knew what combinations of context are more likely to dictate interactions) to benefit the learned context transformation space.

%We identify three rewarding future directions to enhance applications of our model. Developing incremental models for  could enable application to real-time online platforms.  is a promising future direction. Finally, we plan to develop hybrid variants of our approach to accomodate shared users or items and seamlessly incorporate their representations 

%\section{Acknowledgements}
%%\end{document}  % This is where a 'short' article might terminate
%
%\appendix
%
%\section{Headings in Appendices}
%The rules about hierarchical headings discussed above for
%the body of the article are different in the appendices.
%In the \textbf{appendix} environment, the command
%\textbf{section} is used to
%indicate the start of each Appendix, with alphabetic order
%designation (i.e., the first is A, the second B, etc.) and
%a title (if you include one).  So, if you need
%hierarchical structure
%\textit{within} an Appendix, start with \textbf{subsection} as the
%highest level. Here is an outline of the body of this
%document in Appendix-appropriate form:
%
%\begin{acks}
% The authors would like to thank ...
%\end{acks}
\balance
\bibliographystyle{ACM-Reference-Format}
\bibliography{fp0047}

\end{document}

%% file: introduction.tex
%!TEX root = main.tex
\vspace{-7pt}
\section{Introduction}
\label{sec:introduction}

% Recommender systems are a ubiquitous part of our lives today, affecting our interactions with products, services, and content across varied domains. The rapid proliferation of users and items in the social web has aggravated the grey-sheep user/long-tail item challenge in recommendation. Traditional collaborative filtering, as well as the more scalable neural collaborative filtering (NCF) approaches~\cite{vaecf, ncf} continue to suffer from sparse interaction data. In fact, sparsity challenges have become more pronounced \cite{longtail-neural}, motivating us to \textit{learn-to-learn neural embeddings}, i.e., 

The focus of this paper is to learn to build expressive neural collaborative representations of users and items with sparse interaction data. The problem is essential: neural recommender systems are crucial to suggest useful products, services, and content to users online. Sparsity, or the long tail of user interaction, remains a central challenge to traditional collaborative filtering, as well as new neural collaborative filtering (NCF) approaches~\cite{ncf}. Sparsity challenges have become pronounced in neural models~\cite{longtail-neural} owing to generalization and overfitting challenges, motivating us to \textit{learn-to-learn} effective embedding spaces in such a scenario.
Cross-domain transfer learning is a well-studied paradigm to address sparsity in recommendation. However, how recommendation domains are defined plays a key role in deciding the algorithmic challenges. In the most common pairwise cross-domain setting, we can employ cross-domain co-clustering via shared users or items~\cite{wsdm19, ijcai17}, latent structure alignment~\cite{ecmlpkdd13}, or hybrid approaches using both~\cite{www19tmh, www19cngan}. However, recommendation domains with limited user-item overlap are pervasive in real-world applications, such as geographic regions with disparities in data quality and volume (e.g., restaurant recommendation in cities vs. sparse towns). Historically, there is limited work towards such a \textit{few-dense-source}, \textit{multiple-sparse-target} setting, where entity overlap approaches are ineffective. Further, sharing user data entails privacy concerns~\cite{ncf-privacy}.

Simultaneously, context-aware recommendation has become an effective alternative to traditional methods owing to the extensive multi-modal feedback from online users~\cite{cikm18context}. Combinations of contextual predicates prove critical in \textit{learning-to-organize} the user and item latent spaces in recommendation settings. For instance, an \textit{Italian wine restaurant} is a good recommendation for a \textit{high spending} user on a \textit{weekend evening}. However, it is a poor choice for a \textit{Monday afternoon}, when the user is at work. The intersection of restaurant type (an attribute), historical patterns (historical context), and interaction time (interaction context) jointly describe the likelihood of this interaction. 
Our key intuition is to infer such \textit{behavioral invariants} from a \textit{dense-source} domain where we have ample interaction histories of users with wine restaurants and apply (or adapt) these learned invariants to improve inference in \textit{sparse-target} domains. Clustering users who interact under covariant combinations of contextual predicates in different domains lets us better incorporate their behavioral similarities, and analogously, for the item sets as well. The user and item representations in sparse domains can be significantly improved when we combine these transferrable covariances.

Guiding neural representations is also a central theme in gradient-based meta-learning. Recent work~\cite{maml, meta-learn-similar} measures the plasticity of a base-learner via gradient feedback for few-shot adaptation to multiple semantically similar tasks. However, the base-learner is often constrained to simpler architectures (such as shallow neural networks) to prevent overfitting~\cite{DBLP:journals/corr/abs-1812-02391} and requires multi-task gradient feedback at training time~\cite{maml}. This strategy does not scale to the embedding learning problem in NCF, especially in the \textit{many sparse-target} setting. 

Instead, we propose to incorporate the core strengths of meta-learning and transfer learning by defining transferrable neural layers (or meta-layers) via contextual predicates, working in tandem with and guiding domain-specific representations. Further, we develop a novel adaptation approach via regularized residual learning to incorporate new target domains with minimal overheads. Only residual layers and user/item embeddings are learned in each domain while transferring meta-layers, thus also limiting sparse domain overfit. In summary, we make the following contributions:

\begin{description}[leftmargin=0pt, labelindent=\parindent]
%\begin{itemize}
    \item[Contextual invariants for disjoint domains:] We identify the shared task of \textit{learning-to-learn} NCF embeddings via cross-domain contextual invariances. We develop a novel class of pooled contextual predicates to learn descriptive representations in sparse recommendation domains without sharing users or items.
    
%     While invariants are intuitive and well-defined for computer vision tasks, it is harder to unify latent collaborative representations across recommendation domains.
    \item[Tackling the \textit{one-dense}, \textit{many-sparse} scenario:] Our model infers invariant contextual associations via user-item interactions in the dense source domain. Unlike gradient-based meta-learning, we do not sample all domains at train time. We show that it suffices to transfer the source layers to new target domains with an inexpensive and effective residual adaptation strategy. 
    \item[Modular Architecture for Reuse:] Contextual invariants describing user-item interactions are geographically and temporally invariant. Thus we can reuse our meta-layers while only updating the user and item spaces with new data, unlike black-box gradient strategies~\cite{maml}. This also lets us embed new users and items without retraining the model from scratch.
    \item[Strong Experimental Results:] We demonstrate strong experimental results with transfer between dense and sparse recommendation domains in three different datasets - % two publicly available datasets 
    (Yelp Challenge Dataset\footnote{\url{https://www.yelp.com/dataset/challenge}}, Google Local Reviews\footnote{\url{http://cseweb.ucsd.edu/~jmcauley/datasets.html}}) for benchmarking purposes and a large financial transaction dataset from Visa, a major global payments technology company. 
    %massive corporate dataset of financial transactions across multiple regions in the United States. 
    We demonstrate performance and scalability gains on multiple sparse target regions with low interaction volumes and densities by leveraging a single dense source region. 
    %user or item overlaps constraints.
%end{itemize}
\end{description}
%Our work is particularly significant in comparison to previous transfer-learning efforts since it enables adaptation to several sparse target domains with a single dense source-learned meta-model while tackling the constraints of pure gradient-based meta-learning.
 We now summarize related work, formalize our problem, describe our approach, and evaluate the proposed framework.

%% file: related.tex
\section{Related Work}
\label{sec:Related Work}
We briefly summarize a few related lines of work that apply to the sparse inference problem in recommendation: 

%\begin{itemize}
%\item 
\noindent \textbf{Sparsity-Aware Cross-Domain Transfer:} Structure transfer methods regularize the user and item subspaces via principal components~\cite{codebook, aaai10}, joint factorization~\cite{joint-factor1, joint-factor2}, shared and domain-specific cluster structure~\cite{ecmlpkdd13, www19cngan} or combining prediction tasks~\cite{infvae, socialrec} to map user-item preference manifolds. They explicitly map correlated cluster structures in the subspaces. Instead, co-clustering methods use user or item overlaps as anchors for sparse domain inference~\cite{ijcai17, wsdm19}, or auxiliary data~\cite{volkovs2017content, fema} or both~\cite{www19tmh}. It is hard to quantify the volume of users/items or shared content for effective transfer. Further, both overlap-based methods and pairwise structure transfer do not scale to \textit{many sparse-targets}.
% It also scales poorly to non-pairwise transfer settings.
    
%\item 
\noindent \textbf{Neural Layer Adaptation: } A wide-array of layer-transfer and adaptation techniques use convolutional invariants on semantically related images~\cite{joint-adaptation-nets, layer-transfer-coral} and graphs~\cite{motif}. However, unlike convolutional nets, latent collaborative representations are neither interpretable nor permutation invariant~\cite{ncf,vaecf}. Thus it is much harder to establish principled layer-transfer methods for recommendation. We develop our model architecture via novel contextual invariants to enable cross-domain layer-transfer and adaptation.
    
%\item 
\noindent \textbf{Meta-Learning in Recommendation: }
Prior work has considered algorithm selection \cite{federated-meta}, hyper-parameter initialization~\cite{metaparam, kdd19}, shared scoring functions across users~\cite{nips17} and meta-curriculums to train models on related tasks~\cite{kdd19, melu}. Across these threads, the primary challenge is scalability in the multi-domain setting. Although generalizable, they train separate models (over users in ~\cite{nips17}), which can be avoided by adapting or sharing relevant components.

%% file: problemdef.tex
\vspace{-4pt}
\section{Problem Definition}
%\label{sec:Problem Formulation}
%
%In this subsection, we establish a few relevant preliminaries and formalize our context-based cross-domain recommendation setting.
%
%\subsection{Preliminaries}
Consider recommendation domains $\mathbb{D} = \{\mathbf{D}_i\}$ where each $\mathbf{D}_i$ is a tuple $\{\mathcal{U}_{\mathbf{D}_i}, \mathcal{V}_{\mathbf{D}_i}, \mathcal{T}_{\mathbf{D}_i}\}$, with $\mathcal{U}_{\mathbf{D}_i}$, $\mathcal{V}_{\mathbf{D}_i}$ denoting the user and item sets of $\mathbf{D}_i$, and interactions $\mathcal{T}_{\mathbf{D}_i}$ between them. There is no overlap between the user and item sets of any two domains $\mathbf{D}_i$, $\mathbf{D}_j$.

In the implicit feedback setting, each interaction $t \in \mathcal{T}_{\mathbf{D}_i}$ is a tuple $t = (u, \mathbf{c}, v)$ where $u \in \mathcal{U}_{\mathbf{D}_i}, v \in \mathcal{V}_{\mathbf{D}_i}$ and context vector $\mathbf{c} \in \mathbb{R}^{|\mathbf{C}|}$. For the explicit feedback setting, $\mathcal{T}_{\mathbf{D}_i}$  is replaced by ratings $\mathcal{R}_{\mathbf{D}_i}$, where each rating is a tuple $r = (u, \mathbf{c}, v, r_{uv})$, with the rating value $r_{uv}$ (other notations are the same). For simplicity, all interactions in all domains have the same set of context features. In our datasets, the context feature set $\mathbf{C}$ contains three different types of context features, interactional features $\mathbf{C}_{\mathbf{I}}$ (such as time of interaction), historical features $\mathbf{C}_{\mathbf{H}}$ (such as a user's average spend), and attributional features $\mathbf{C}_{\mathbf{A}}$ (such as restaurant cuisine or user age). Thus each context vector $\mathbf{c}$ contains these three types of features for that interaction, i.e., $\mathbf{c} = [\mathbf{c}_{\mathbf{I}}, \mathbf{c}_{\mathbf{H}}, \mathbf{c}_{\mathbf{A}}]$. 

Under implicit feedback, we rank items $v \in \mathcal{V}_{\mathbf{D}}$ given user $u \in \mathcal{U}_{\mathbf{D}}$ and context $\mathbf{c}$.  In the explicit feedback scenario, we predict rating $r_{uv}$ for $v \in \mathcal{V}_{\mathbf{D}}$ given $u \in \mathcal{U}_{\mathbf{D}}$ and $\mathbf{c}$. Our transfer objective is to reduce the rating or ranking error in a set of disjoint sparse target domains $\{\mathbf{D}_t\} \subset \mathbb{D}$ given the dense source domain $\mathbf{D}_s \in \mathbb{D}$.

%% file: model.tex
\section{Our Approach}
\label{sec:Model Description}

In this section, we describe a scalable, modular architecture to extract pooled contextual invariants and employ them to guide the learned user and item embedding spaces.

%In this section, we formalize cross-domain behavioral invariants via pooled contextual features and describe a scalable, modular architecture to extract and integrate them with the user and item embedding spaces. We predict likely interactions with the resulting context conditioned representations. Our formalization provides an automatic separation between the domain-specific embeddings and shared transformations grounded on context (meta). Subsequently, we develop meta transfer algorithms to adapt the shared components to sparse target domains in \Cref{subsec:Meta-Transfer}. 
\vspace{-4pt}
\subsection{Modular Architecture}
\label{subsec:Meta-Problem Formulation}

We achieve context-guided embedding learning via four synchronized neural modules with complementary semantic objectives:
%Our crucial intuition is the presence of behavioral invariants at the intersection of multiple contextual indicators, i.e., changing a single feature could drastically alter the likelihood of interaction, even with all other features held constant. Additive or linear models struggle to learn interaction effects \cite{nfm, mult-google}. Inspired by the past success of low-rank pooling models \cite{bilinear, mult-google}, we propose a low-rank multi-linear context representation to capture behavioral invariants.

\begin{figure}[t]
    \centering
    \vspace{-4pt}
    \caption{Our overall recommender architecture, highlighting all four modules, $\mathcal{M}^1 \text{ to } \mathcal{M}^4$}
    \vspace{-3pt}
    \includegraphics[clip, trim=1cm 0.1cm 17cm 3.5cm, width=0.82\linewidth]{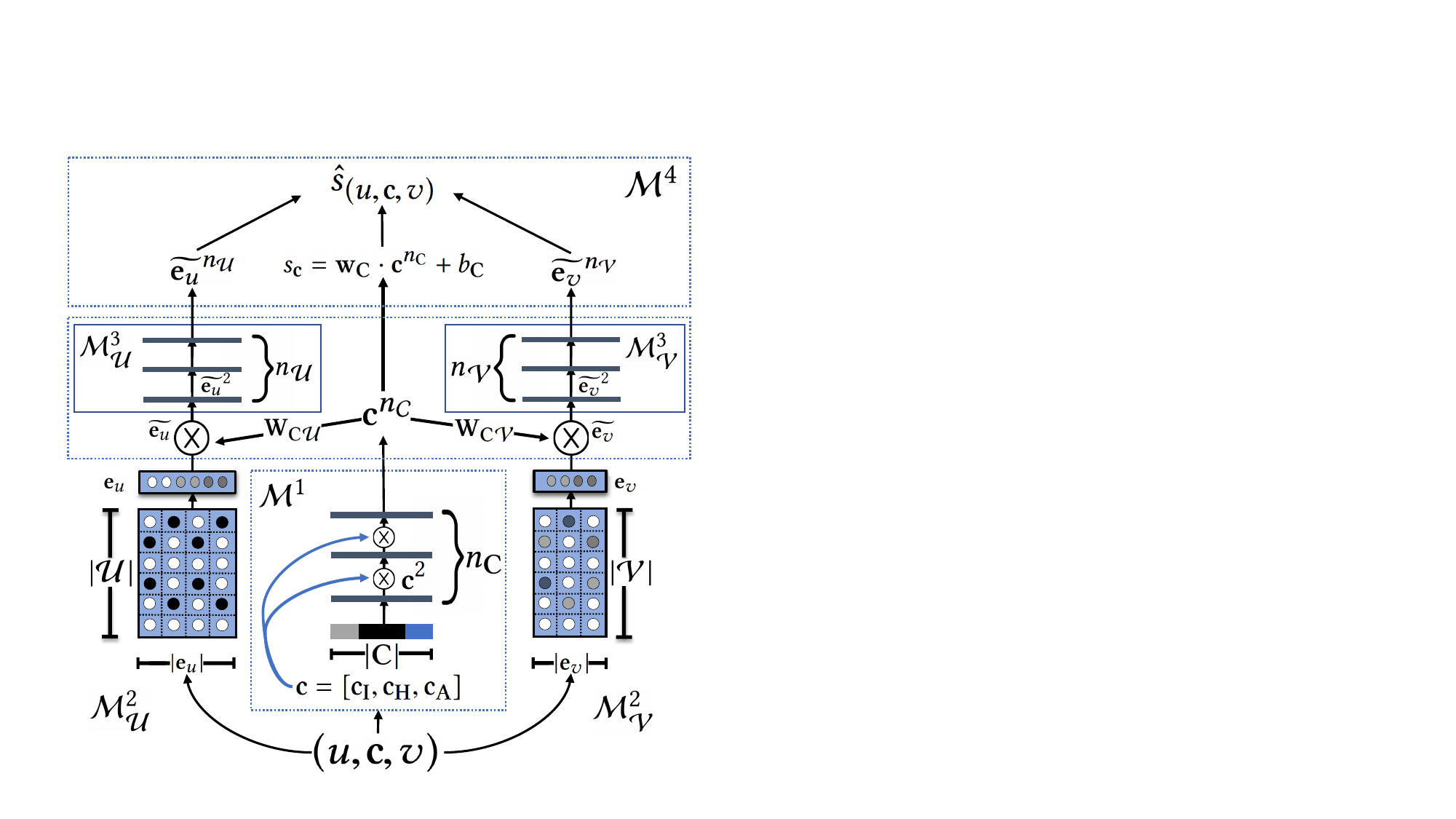}
    \vspace{-28pt}
    \label{fig:neural_1}
    \end{figure}

%We leverage the contextual features of user interactions to describe these behavior patterns and learn representative user and item embedding spaces. This forms the core of our \textit{learn-to-learn} formulation, whereby we learn invariants on the dense source domain where they manifest in the observed user-item interactions. Critical invariants often appear at the intersection of multiple contextual indicators, e.g., changing a single context feature such as time of the day could drastically alter the likelihood of interaction even with other features held constant. Additive models struggle to learn interaction effects \cite{nfm}. Past work has even shown deep neural networks driven by linear transforms struggle to infer pooled or multiplicative factors \cite{mult-google}. Inspired by the past success of low-rank pooling models \cite{bilinear, mult-google}, we propose a multi-linear low rank pooled representation to capture invariant context transforms describing user behavior. Our architecture broadly consists of four components:
\begin{itemize}
\item \textbf{Context Module $\mathcal{M}^{1}$: } Extracts contextual invariants driving user-item interactions in the dense source domain.
\item \textbf{Embedding Modules $\mathcal{M}^{2}_{\mathcal{U}}, \mathcal{M}^{2}_{\mathcal{V}}$: } Domain-specific user and item embedding spaces ($\mathcal{U}$, $\mathcal{V}$ denote users and items).
\item \textbf{Context-conditioned Clustering Modules $\mathcal{M}^{3}_{\mathcal{U}}, \mathcal{M}^{3}_{\mathcal{V}}$:\\} $\mathcal{M}^{3}_{\mathcal{U}}$ and $\mathcal{M}^{3}_{\mathcal{V}}$ reorient the user and item embeddings with the contextual invariants extracted by $\mathcal{M}^{1}$ respectively.
\item \textbf{Mapping/Ranking Module $\mathcal{M}^{4}$: } Generate interaction likelihoods with the context-conditioned representations of $\mathcal{M}^{3}$.
\end{itemize}
Context-driven modules $\mathcal{M}^{1}$, $\mathcal{M}^{3}$, and $\mathcal{M}^{4}$ contain the meta-layers that are transferred from the dense to the sparse domains (i.e., shared or meta-modules), while $\mathcal{M}^{2}$ contains the domain-specific user and item representations. Our architecture provides a separation between the domain-specific $\mathcal{M}^{2}$ module and shared context-based transforms in the other modules (\Cref{fig:neural_1}).

%\noindent \textbf{The importance of low-rank pooling: } We extract the most informative contextual combinations to describe each interaction. Specially, the output of the context transform component is composed of \textit{n}-variate combinations of the contextual features. We enable data-driven selection of pooled \textit{n}-variate factors to address combinatorial explosion with growing context. Further, a tiny proportion of pooled combinations are expected to play a crucial role in the user-item interactions. Our approach facilitates their discovery by adaptive weighting the chosen set of multi-linear factors.
\vspace{-4pt}
\subsection{Module Description}
\label{subsec:Arch}
We now detail each module in our overall architecture.
%\begin{itemize}
%\item 

\subsubsection{Context Module $\mathcal{M}^1$} 

User-item interactions are driven by context feature intersections that are inherently \textit{multiplicative} (i.e., assumptions of independent feature contributions are insufficient), and are often missed in the Naive-Bayes assumption of additive models such as feature attention~\cite{nfm, mult-google}. Inspired by the past success of low-rank feature pooling~\cite{bilinear, mult-google}, our context module extracts low-rank multi-linear combinations of context to describe interactions and build expressive representations.
   % $$\mathcal{M}^1$ contains $n$ stacked layers, with each layer transforming an interaction $(u, \mathbf{c}, v)$ as follows$
The first layer in $\mathcal{M}^1$ transforms context $\mathbf{c}$ of an interaction $(u, \mathbf{c}, v)$ as follows: 
    \begin{equation}
    \mathbf{c}^{2} = \sigma(\underbrace{\mathbf{W}^2\mathbf{c} \oplus (\mathbf{b}^{2} \otimes \mathbf{c})}_{\text{Weighted linear transform}}) \hspace{10pt} \otimes \hspace{-20pt} \underbrace{ \mathbf{c}}_{\text{Element-wise interaction}}
    \end{equation}
   where $\oplus, \otimes$ denote element-wise product and sum, i.e.,
    \begin{equation}
    \mathbf{c}^{2}_{i} \propto \mathbf{c}_i \times \sigma(\mathbf{b}^2_{i}\mathbf{c}_i  + \sum_{j}\mathbf{W}^2_{ij} \mathbf{c}_j) 
    \vspace{-8pt}
    \end{equation}  
    Thus, $\mathbf{c}^{2}_{i}$ ($i^{th}\text{-component}$ of $\mathbf{c}^{2}$) incorporates a weighted bivariate interaction between $\mathbf{c}_{i}$ and other context factors $\mathbf{c}_j$, including itself. We then repeat this transformation over multiple stacked layers with each layer using the previous output:
	\begin{equation}    
    \mathbf{c}^{n} = \sigma(\mathbf{W}^{n}\mathbf{c}^{n-1} \oplus (\mathbf{b}^{n} \otimes \mathbf{c}^{n-1})) \otimes \mathbf{c}
    \label{cn}
    \end{equation}
 Each layer interacts \textit{n-variate} terms from the previous layer with $\mathbf{c}$ to form \textit{n+1-variate} terms. However, since each layer has only $|\mathbf{C}|$ outputs (i.e., low-rank), $\mathbf{W}^{n}$ prioritizes the most effective \textit{n-variate} combinations of $\mathbf{c}$ (typically, a very small fraction of all combinations is useful). We can choose the number of layers $n_{\mathbf{C}}$ depending on the required order of the final combinations $\mathbf{c}^{n_{\mathcal{C}}}$.
%     $$\mathbf({c}_{2})_{i} = \mathbf{c}_{i} \times \sum_{j = 1}^{|\mathbf{c}|}\mathbf{W}_{i,j}^{1} \mathbf{c}_{j} = \sum_{j = 1}^{|\mathbf{c}|}\mathbf{W}_{i,j}^{1} \mathbf{c}_{i}\mathbf{c}_{j}$$

    \textit{Multimodal Residuals for Discriminative Correlation Mining:}
  In addition to discovering the most important context combinations, we incorporate the information gain associated with pairwise interactions of context features \cite{featurepool}. For instance, the item cost feature is more informative in interactions where users deviate from their historical spending patterns. Specifically, pairs of signals (e.g., cost \& user history) enhance or diminish each other's impact, i.e.,
  \begin{equation}
   \mathbf{c}_{i} = \mathbf{c}_{i} + \Sigma_{j}\delta_{\mathbf{c}_i | \mathbf{c}_j}
   \label{mm1}
   \end{equation}
     We simplify \Cref{mm1} by only considering cross-modal effects across interactional, historical and attributional context, i.e., 
\begin{equation}
\delta_{\mathbf{c}_{\mathbf{I}}|\mathbf{c}_{\mathbf{H}}, \mathbf{c}_{\mathbf{A}}} = \underbrace{\mathbf{s_{\mathbf{I}}}}_{\text{Scaling factor}}  \hspace{-5pt}\otimes\hspace{7pt} \underbrace{\textit{tanh}(\mathbf{W}_{\mathbf{IH}} \times \mathbf{c}_{\mathbf{H}} + \mathbf{W}_{\mathbf{IA}} \times \mathbf{c}_{\mathbf{A}} + \mathbf{b}_{\mathbf{I}})}_{\text{Info gain/loss}}
\label{multimodal}
\end{equation}
 and likewise for $\delta_{\mathbf{c}_{\mathbf{H}}}$, $\delta_{\mathbf{c}_{\mathbf{A}}}$. Information gains are computed before $\mathbf{c}^{2}$ to cascade to further layers.

%    Note that we learn a scaling parameter $\mathbf{s}$ and weight $\mathbf{W}$  for each context mode. The above residual transforms are applied to raw context $\mathbf{c}$ before the first transformation layer to enable a cascading effect over the other layers.

% \noindent \textbf{Organize the User Embedding Space: }    

\begin{table}[t]
 \centering
 \caption{Modules and Learned Parameter Notations}
 \vspace{-12pt}
 \begin{tabular}{@{}p{0.32\linewidth}p{0.66\linewidth}@{}}
  \toprule
\textbf{Modules} & \textbf{Learned Parameters}\\  
  \midrule
     \multirow{1}{*}{\textbf{Domain-Specific} } &  \textbf{Embeddings} $\mathbf{e}_u \forall u \in \mathcal{U}_{\mathbf{D}}$, $\mathbf{e}_{v} \forall v \in \mathcal{V}_{\mathbf{D}}$\\
     ($\mathcal{M}^{2}_{\mathcal{U}}, \mathcal{M}^{2}_{\mathcal{V}}$) & \textbf{Biases} (only under explicit feedback) $s, s_u  \forall u \in \mathcal{U}_{\mathbf{D}}, s_{v} \forall v \in \mathcal{V}_{\mathbf{D}}$ \\
     \midrule
     
  \multirow{2}{*}{\textbf{Shared Modules}} &  $\mathcal{M}^{1}$ \hspace{2pt} \cref{cn} \hspace{3pt} $(\mathbf{W}^{i}, \mathbf{b}^{i}) \forall i = [1, \cdots , n_{\mathbf{C}}]$ \\
  
 &  $\mathcal{M}^{1}$ \hspace{2pt} \cref{multimodal} \hspace{3pt} $\mathbf{s}_{\mathbf{I}},  \mathbf{s}_{\mathbf{H}},  \mathbf{s}_{\mathbf{A}}; \mathbf{W}_{\mathbf{I}},  \mathbf{W}_{\mathbf{H}},  \mathbf{W}_{\mathbf{A}}$ \\
 \multirow{2}{*}{($\mathcal{M}^{1}$, $\mathcal{M}^{3}, \mathcal{M}^4$)} & $\mathcal{M}^{3}$ \hspace{2pt} \cref{link} \hspace{3pt} $\mathbf{W}_{\mathbf{C}\mathcal{U}}, \mathbf{W}_{\mathbf{C}\mathcal{V}}$\\
 
 & $\mathcal{M}^{3}_{\mathcal{U}}$ \cref{eq:ccr} \hspace{3pt}  $(\mathbf{W}^{i}_{\mathcal{U}}, \mathbf{b}^{i}_{\mathcal{U}}) \forall i = [1, \cdots , n_{\mathcal{U}}]$ \\
 
 & $\mathcal{M}^{3}_{\mathcal{V}}$ \cref{eq:ccr} \hspace{3pt}  $(\mathbf{W}^{i}_{\mathcal{V}}, \mathbf{b}^{i}_{\mathcal{V}}) \forall i = [1, \cdots , n_{\mathcal{V}}]$\\
 
 & $\mathcal{M}^{4}$ \hspace{2pt} \cref{sc} \hspace{3pt} $\mathbf{W}_{\mathbf{C}}, \mathbf{b}_{\mathbf{C}}$\\
  \bottomrule
 \end{tabular}
 \label{notation-table}
 \vspace{-14pt}
\end{table}

\subsubsection{Context Conditioned Clustering $\mathcal{M}^3$}
We combine domain-specific embeddings $\mathcal{M}^2$ with the context combinations extracted by $\mathcal{M}^1$ to generate context-conditioned user and item representations. Specifically, we introduce the following bilinear transforms,
 \begin{align}
    \widetilde{\mathbf{e}_{u}} = \mathbf{e}_{u} \otimes  \sigma(\mathbf{W}_{\mathbf{C}\mathcal{U}} \times \mathbf{c}^{n_\mathbf{C}}) \\
    \widetilde{\mathbf{e}_{v}} = \mathbf{e}_{v} \otimes  \sigma(\mathbf{W}_{\mathbf{C}\mathcal{V}} \times \mathbf{c}^{n_\mathbf{C}})
    \label{link}
   \end{align}

%To achieve this organization of the embeddings, we backpropagate the extracted multi-linear context embeddings $\mathbf{c}^{n_\mathbf{C}}$ into the user embedding space and create context conditioned clusters of users for item ranking.

%   The user embedding space, $\mathbf{e}_u, u \in \mathcal{U}_{D}$, is organized to reflect the contextual preferences of users. The same motivation holds good for the item embedding space as well. 

%    The bilinear layers eliminate the irrelevant dimensions of the user and item embeddings to generate the conditioned representations $\widetilde{\mathbf{e}_{u}}$ and $\widetilde{\mathbf{e}_{v}}$. Further, bilinear transforms enable cross-domain \textit{dimensional-uniformity} of embeddings. Context features are analogously transformed and backpropagated into their user and item spaces.
    
%     The vectors $\widetilde{\mathbf{e}_{u}}$ and $\widetilde{\mathbf{e}_{i}}$ represent the context based clusters of users and items, based upon the $\mathit{n}$-variate context combinations in $\mathbf{c}^{n_\mathbf{C}}$.
    
%    \noindent $\textbf{D.}$ \textbf{Cluster Mapping - } 
    
where, $\mathbf{W}_{\mathbf{C}\mathcal{U}} \in \mathbb{R}^{{|\mathbf{e}_u| \times |\mathbf{C}}|}$, $\mathbf{W}_{\mathbf{C}\mathcal{V}} \in \mathbb{R}^{{|\mathbf{e}_v| \times |\mathbf{C}}|}$ are learned parameters that map the most relevant context combinations to the user and item embeddings. We further introduce $n_\mathcal{U}$ feedforward \textit{RelU} layers to cluster the representations,
\begin{align}
    \widetilde{\mathbf{e}_{u}}^{2} = \sigma(\mathbf{W}_{\mathcal{U}}^{2}\widetilde{\mathbf{e}_{u}} + \mathbf{b}_{\mathcal{U}}^{2})\\
    \widetilde{\mathbf{e}_{u}}^{n} = \sigma(\mathbf{W}_{\mathcal{U}}^{n}\widetilde{\mathbf{e}_{u}}^{n-1} + \mathbf{b}_{\mathcal{U}}^{n})
    \label{eq:ccr}
    \end{align}
Analogously, we obtain context-conditioned item representations $\widetilde{\mathbf{e}_{v}}^{2}, \cdots, \widetilde{\mathbf{e}_{v}}^{n_\mathcal{V}}$ with $n_\mathcal{V}$ feedforward \textit{RelU} layers. 

The bilinear transforms in \cref{link} introduce \textit{dimension alignment} for both $\widetilde{\mathbf{e}_{u}}^{n_\mathcal{U}}$ and $\widetilde{\mathbf{e}_{v}}^{n_\mathcal{V}}$ with the context output $\mathbf{c}^{n_{\mathbf{C}}}$. Thus, when $\mathcal{M}^3$ and $\mathcal{M}^1$ layers are transferred to a sparse target domain, we can directly backpropagate to guide the target domain user and item embeddings with the target domain interactions.

%    The score for $u, v$ under context $\mathbf{c}$ (i.e., module $\mathcal{M}^4$) is reduced to just the dot product:
%        $$\hat{s}_{u, \mathbf{c}, v} = \widetilde{\mathbf{e}_{u}}^{n_\mathcal{U}} \cdot \widetilde{\mathbf{e}_{v}}^{n_\mathcal{V}}$$
%However in practice, the above loss function results in uninteresting low-variance samples dominating the learning process, resulting in slower convergence, less novelty and inaccurate user representations. In the next subsection, we tackle this challenge.

 \vspace{-3pt}
\subsection{Source Domain Training Algorithm}    
\label{subsec:Train}
In the source domain, we train all modules and parameters (\Cref{notation-table}) with ADAM optimization~\cite{adam} and dropout regularization~\cite{dropout}.

\subsubsection{Self-paced Curriculum via Contextual Novelty}
Focusing on harder data samples accelerates and stabilizes stochastic gradients \cite{hardsample1, active_bias}. Since our learning process is grounded on context, novel interactions display uncommon or \textit{interesting} context combinations. Let $\mathcal{L}_{(u, \mathbf{c}, v)}$ denote the loss function for an interaction $(u, \mathbf{c}, v)$. We propose an inverse novelty measure referred as the context-bias, $s_{\mathbf{c}}$, which is self-paced by the context combinations learned by $\mathcal{M}^1$ in \Cref{cn},
\begin{equation} 
   s_{\mathbf{c}} = \mathbf{w}_{\mathbf{C}} \cdot \mathbf{c}^{n_\mathbf{C}} + b_{\mathbf{C}}
   \label{sc}
\end{equation}
We then attenuate the loss $\mathcal{L}_{(u, \mathbf{c}, v)}$ for this interaction as,
\begin{equation}
	 \mathcal{L}_{(u, \mathbf{c}, v)}^{\mathbf{'}} =  \mathcal{L}_{(u, \mathbf{c}, v)} - s_{\mathbf{c}}
	 \label{eq:context_bias}
\end{equation}
The resulting novelty loss $\mathcal{L}_{(u, \mathbf{c}, v)}^{\mathbf{'}}$ decorrelates interactions \cite{decorrelate, var_reduction} by emulating variance-reduction in the \textit{n-variate} pooled space of $\mathbf{c}^{n_\mathbf{C}}$. $\mathcal{L}_{(u, \mathbf{c}, v)}^{\mathbf{'}}$ determines the user and item embedding spaces, inducing a novelty-weighted training curriculum focused on harder samples as training proceeds. We now describe loss $\mathcal{L}_{(u, \mathbf{c}, v)}$ for the explicit and implicit feedback scenarios.

%The novelty factor is self-paced by the $\mathbf{c}^{n_\mathbf{C}}$ transform generated by module $\mathcal{M}^1$. 
% 
%A practical example could be users visiting inexpensive restaurants in proximity to their common locations. Such examples can be predicted from context alone, and thus exhibit a stronger variance These examples constitute a large proportion of the training samples; fewer examples exhibit novel or diverse interests of users and the corresponding context. 

%\noindent $\textbf{E.}$ \textbf{Learn the Context Bias Component - } Finally,  The key intuition for this term is that some contexts are universally more likely to result in a restaurant visit in comparison to others. For instance, the bias to visit a highly rated restaurant in proximity to the user is expected to be significantly more than that of an average restaurant that is inaccessible to the user. To obtain this context bias score, we learn a simple dot-product layer,
%    $$s_{\mathbf{c}} = \mathbf{w}_{\mathbf{c}} \cdot \mathbf{c}^{n_\mathbf{C}} + b_{\mathbf{c}}$$

\subsubsection{Ranking our Recommendations}
In the \textit{implicit feedback setting}, predicted likelihood $\hat{s}_{(u, \mathbf{c}, v)}$ is computed with the context-conditioned embeddings (\Cref{eq:ccr}) and context-bias (\Cref{eq:context_bias}) as,
\begin{equation}
 \hat{s}_{(u, \mathbf{c}, v)} = \widetilde{\mathbf{e}_{u}}^{n_\mathcal{U}} \cdot \widetilde{\mathbf{e}_{v}}^{n_\mathcal{V}} + s_{\mathbf{c}}
 \end{equation}
 The loss for all the possible user-item-context combinations in domain $\mathbf{D}$ is,
 \begin{equation}
 \mathcal{L}_{\mathbf{D}} = \sum_{u \in \mathcal{U}_{\mathbf{D}}} \sum_{v \in \mathcal{V}_{\mathbf{D}}} \sum_{\mathbf{c} \in \mathbb{R}^{|\mathbf{c}|}} || \mathbb{I}_{(u, \mathbf{c}, v)} - \hat{s}_{(u, \mathbf{c}, v)} ||^{2}
 \label{basic_loss}
 \end{equation}
        where $\mathbb{I}$ is the binary indicator $(u, \mathbf{c}, v) \in \mathcal{T}_{\mathbf{D}}$. $\mathcal{L}_{\mathbf{D}}$ is intractable due to the large number of contexts $\mathbf{c} \in \mathbb{R}^{|\mathbf{c}|}$. We develop a negative sampling approximation for implicit feedback with two learning objectives - identify the likely item given the user and interaction context, and identify the likely context given the user and the item. We thus construct two negative samples for each $(u, \mathbf{c}, v) \in \mathcal{T}_{\mathbf{D}}$ at random: Item negative with the true context, $(u, \mathbf{c}, v^{-})$ and context negative with the true item, $(u, \mathbf{c}^{-}, v)$. $\mathcal{L}_{\mathbf{D}}$ then simplifies to,
\begin{equation}        
        \mathcal{L}_{\mathbf{D}} = \sum_{\mathcal{T}_{\mathbf{D}}} || 1 - \hat{s}_{(u, \mathbf{c}, v)} ||^{2} + \sum_{(u, \mathbf{c}, v^{-})} ||\hat{s}_{(u, \mathbf{c}, v^{-})} || + \sum_{(u, \mathbf{c}^{-}, v)} ||\hat{s}_{(u, \mathbf{c}^{-}, v)} || 
        \label{negsample_loss}
\end{equation}
        
\textit{In the explcit feedback setting}, we introduce two additional bias terms, one for each user, $s_{u}$ and one for each item, $s_v$. These terms account for user and item rating eccentricities (e.g., users who always rate well), so that the embeddings are updated with the relative rating differences. Finally, global bias $s$ accounts for the rating scale, e.g., 0-5 vs. 0-10. Thus the predicted rating is given as,

        \begin{equation}
        \hat{r}_{(u, \mathbf{c}, v)} = \widetilde{\mathbf{e}_{v}}^{n_\mathcal{V}} \cdot \widetilde{\mathbf{e}_{u}}^{n_\mathcal{U}} + s_{\mathbf{c}} + s_{u} + s_{v} + s
        \label{explicit_rating}
        \end{equation}
        Negative samples are not required in the explicit feedback setting,
        
        \begin{equation}
        \mathcal{L}_{\mathbf{D}}^{explicit} = \sum_{(u, \mathbf{c}, v, r_{uv}) \in \mathcal{R}_{\mathbf{D}}}  || r_{uv} - \hat{r}_{(u, \mathbf{c}, v)} ||^{2}
        \label{explicit_loss}
        \end{equation}
    
We now detail our approach to transfer the shared modules from the source domain to sparse target domains.

%% file: model_adaptation.tex
\section{Transfer to Target Domains}
\label{subsec:Meta-Transfer}
Our formulation enables us to train the shared modules $(\mathcal{M}^1)_{\mathbf{S}}, (\mathcal{M}^3)_{\mathbf{S}}$ and $(\mathcal{M}^4)_{\mathbf{S}}$ on a dense source domain $\mathbf{S}$, and transfer them to a sparse target domain $\mathbf{T}$ to guide its embedding module $(\mathcal{M}^2)_{\mathbf{T}}$. Each shared module $\mathcal{M}$ encodes inputs $\mathbf{x}_{\mathcal{M}}$ to generate output representations $\mathbf{y}_{\mathcal{M}}$. In each domain $\mathbf{T}$, module $(\mathcal{M})_{\mathbf{T}}$ determines the joint input-output distribution, 
\begin{equation}
p_{\mathbf{T}}(\mathbf{y}_{\mathcal{M}}, \mathbf{x}_{\mathcal{M}}) = p_{\mathbf{T}}(\mathbf{y}_{\mathcal{M}} | \mathbf{x}_{\mathcal{M}}) \times p_{\mathbf{T}}(\mathbf{x}_{\mathcal{M}})
\label{joint}
\end{equation}
where the parameters of $(\mathcal{M})_{\mathbf{T}}$ determine the conditional $p_{\mathbf{T}}(\mathbf{y}_{\mathcal{M}} | \mathbf{x}_{\mathcal{M}})$ and $p_{\mathbf{T}}(\mathbf{x}_{\mathcal{M}})$ describes the inputs to module $(\mathcal{M})_{\mathbf{T}}$ in domain ${\mathbf{T}}$.
\textbf{Adaptation: }There are two broad strategies to adapt module $\mathcal{M}$ to a new target domain ${\mathbf{T}}$:
\begin{itemize}
\item \textbf{Parameter Adaptation:} We can retrain the parameters of module $\mathcal{M}$ for target domain $\mathbf{T}$ thus effectively changing the conditional $p_{\mathbf{T}}(\mathbf{y}_{\mathcal{M}} | \mathbf{x}_{\mathcal{M}})$ in \cref{joint}, or, 
\item \textbf{Input Adaptation:} Modify the input distribution $p_{\mathbf{T}}(\mathbf{x}_{\mathcal{M}})$ in each domain $\mathbf{T}$ without altering the parameters of $\mathcal{M}$.
\end{itemize}
We now explore module transfer with both types of adaptation strategies towards achieving three key objectives. First, the transferred modules must be optimized to be effective on each target domain $\mathbf{T}$. Second, we aim to minimize the computational costs of adapting to new domains by maximizing the reuse of module parameters between the source $\mathbf{S}$ and target domains $\mathbf{T}$. Finally, we must take care to avoid overfitting the transferred modules to the samples in the sparse target domain $\mathbf{T}$.
%\begin{itemize}
%\item \textbf{Target Domain Fit:} 
%\item \textbf{Minimize computational overheads per target:} 
%\item \textbf{Maximize shared parameters:} Maximize direct source-to-target parameter reuse for simplicity and efficiency.
%\item \textbf{Prevent Overfitting:} 
%\end{itemize}
\vspace{-4pt}
\subsection{Direct Layer-Transfer}
\label{subsec:DLT}
We first train all four modules on the source $\mathbf{S}$ and each target domain $\mathbf{T}$ in isolation. We denote these pre-trained modules as $(\mathcal{M}^i)_{\mathbf{S}}$ and $(\mathcal{M}^i)_{\mathbf{T}}$ for source domain $\mathbf{S}$ and a target domains $\mathbf{T}$ respectively. We then replace the shared modules in all the target domain models with the source-trained version, i.e., $(\mathcal{M}^{1})_{\mathbf{T}} = (\mathcal{M}^{1})_{\mathbf{S}}$,  $(\mathcal{M}^{3})_{\mathbf{T}} = (\mathcal{M}^{3})_{\mathbf{S}}$, $(\mathcal{M}^{4})_{\mathbf{T}} = (\mathcal{M}^{4})_{\mathbf{S}}$, while the domain-specific embeddings $(\mathcal{M}^{2})_{\mathbf{T}}$ are not changed in the target domains. Clearly, direct layer-transfer involves no overhead and trivially prevents overfitting. However, we need to adapt the transferred modules for optimal target performance, i.e., either adapt the parameters or the input distributions for the transferred modules in each target $\mathbf{T}$. We now develop these adaptation strategies building on layer-transfer.
%We replicate the domain-independent parameters and meta-modules in \Cref{notation-table} in the target domain, while the domain-specific parameters are pre-trained on the target data before layer-transfer. We can then obtain recommendation with the pre-trained user parameters and the transferred meta-layers.
% While direct layer-transfer has produced impressive results across a range of Computer Vision tasks \cite{layer-transfer-empirical},
%Our key objectives are two-fold - avoid overfitting to the sparse target domain while adapting the transferred layers, and scale laterally, i.e., adapt to new sparse target domains with minimal computational/storage overheads.
%
%While direct layer-transfer is highly scalable; there is significant scope to adapt the layers to the eccentricities of the target domain and optimize performance. We begin by introducing simulated annealing as a feasible adaptation strategy, and proceed to address its scalability concerns. 
%Note that the primary goal in the target domain is to learn representative user and merchant embeddings with a relatively low volume and density of transactional data. The key strength of our approach is to adapt the pre-determined contextual combinations, user and merchant clustering layers and  This enables our model to efficiently leverage the smaller volume of transactional data in the target domain. 
\begin{table*}[t]
 \centering
 \caption{Comparing the objectives in \Cref{subsec:Meta-Transfer} addressed by our meta-transfer approaches for sparse target domains}
 \vspace{-12pt}
 \begin{tabular}{@{}p{0.18\linewidth}p{0.16\linewidth}p{0.155\linewidth}p{0.21\linewidth}p{0.22\linewidth}@{}}
  \toprule
     \textbf{Adaptation Method} & \textbf{Target Adaptation} & \textbf{Resists Overfitting} & \textbf{Extra compute per target} & \textbf{Extra parameters per target}  \\
  \midrule
\textbf{Layer-Transfer} & No adaptation & Yes, trivially & None & None, module params reused\\
%\midrule
\textbf{Simulated Annealing} & Yes, module params & Yes, stochastic & All parameter updates & All module params (~\Cref{notation-table})\\
%\midrule
%\textbf{} & Yes, adapts module inputs & No & Residual layer updates & Residual layers only\\
\textbf{Regularized Residuals} & Yes, module inputs & Yes, via distributional consistency & Residual layer updates with distributional regularization & Residual layer parameters\\
  \bottomrule
 \end{tabular}
 \label{methods}
 \vspace{-12pt}
\end{table*}
\vspace{-4pt}
\subsection{Simulated Annealing}
\label{subsec:anneal}
Simulated annealing is a stochastic local-search algorithm, that implicitly thresholds parameter variations in the gradient space by decaying the gradient learning rates~\cite{anneal}. As a simple and effective adaptation strategy, we anneal each transferred module $\mathcal{M}$ in the target domain $\mathbf{T}$ with exponentially decaying learning rates to stochastically prevent overfitting: 
\begin{equation}
(m)_{b+1} = (m)_{b} + \eta_{b}\frac{\partial\mathcal{L}_{b}}{\partial m}, \hspace{3pt} \eta_{b} = \eta_{0}e^{-\lambda b}
\label{eq:anneal}
\end{equation}
where $m$ denotes any parameter of transferred module $\mathcal{M}$ (\Cref{notation-table}), $b$ is the stochastic gradient batch index in the target domain and $\mathcal{L}_b$ is the batch loss for batch $b$. Our annealing strategy stochastically generates a robust parameter search schedule for transferred modules $\mathcal{M}^{1}, \mathcal{M}^{3}, \mathcal{M}^{4}$, with $\eta_{b}$ decaying to zero after one annealing epoch. While annealing the transferred modules, domain-specific module $\mathcal{M}^2$ is updated with the full learning rate $\eta_{0}$.
Clearly, annealing modifies the conditional $p_{\mathbf{T}}(\mathbf{y}_{\mathcal{M}} | \mathbf{x}_{\mathcal{M}})$ in \cref{joint} via parameter adaptation. However, annealing transferred modules in each target domain is somewhat expensive, and the annealed parameters are not shareable, thus causing scalability limitations in the one-to-many transfer scenario. We now develop a lightweight residual adaptation strategy to achieve input adaptation without modifying any shared module parameters in the target domains to overcome the above scalability challenges.
\vspace{-4pt}
\subsection{Distributionally Regularized Residuals}
\label{subsec:DRR}
% \subsubsection{adapt:anneal}%Adaptation approaches that modify $p(\mathbf{y}_{\mathcal{M}} | \mathbf{x}_{\mathcal{M}})$ must alter module parameters in each new target domain, resulting in scalability challenges. Instead, w
We now develop an approach to reuse the source modules with target-specific input adaptation, thus addressing the scalability concerns of parameter adaptation methods.
\subsubsection{Enabling Module Reuse with Residual Input  Adaptation}
In \cref{joint}, module $\mathcal{M}$ implements the conditional $p(\mathbf{y}_{\mathcal{M}} | \mathbf{x}_{\mathcal{M}})$. To maximize parameter reuse, we share these modules across the source and target domains (i.e., $p_{\mathbf{T}}(\mathbf{y}_{\mathcal{M}} | \mathbf{x}_{\mathcal{M}}) = p_{\mathbf{S}}(\mathbf{y}_{\mathcal{M}} | \mathbf{x}_{\mathcal{M}})$) and introduce target-specific residual perturbations to account for their eccentricities~\cite{classifier-adap} by modifying the input distributions $p_{\mathbf{T}}(\mathbf{x}_{\mathcal{M}})$. Target-specific input adaptation overcomes the need for an expensive end-to-end parameter search. Our adaptation problem thus reduces to learning an input modifier for each target domain $\mathbf{T}$ and shared module $\mathcal{M} \in [\mathcal{M}^1, \mathcal{M}^3, \mathcal{M}^4]$, i.e., for each $\mathcal{M}, \mathbf{T}$.
%\begin{equation}
%x_{\mathcal{M}}^{\mathbf{T}} = f_{\mathcal{M}}^{\mathbf{T}}(x_{\mathcal{M}})
%\end{equation}
Residual transformations enable the flow of information between layers without the gradient attenuation of inserting new non-linear layers, resulting in numerous optimization advantages~\cite{resnet}. Given the module-input $\mathbf{x}_{\mathcal{M}}$ to the shared module $\mathcal{M}$ in target domain $\mathbf{T}$, we learn a module and target specific residual transform:
\begin{equation}
\mathbf{x}_{\mathcal{M}} = \mathbf{x}_{\mathcal{M}} + \delta_{\mathcal{M}, \mathbf{T}}(\mathbf{x}_{\mathcal{M}})
\end{equation}
The form of the residual function $\delta$ is flexible. We chose a single non-linear residual layer, $\delta(\mathbf{x}) = tanh(\mathbf{W}\mathbf{x} + \mathbf{b})$. We can intuitively balance the complexity and number of such residual layers.
Note that the above residual strategy involves learning the $\delta_{\mathcal{M}, \mathbf{T}}$ layers with feedback from only the sparse target domain samples. To avoid overfitting, we need a scalable regularization strategy to regularize $p_{\mathbf{T}}(\mathbf{x}_{M})$ in each target domain. We propose to leverage the source input distribution as a common baseline for all the target domains, i.e., intuitively, $p_{\mathbf{S}}(\mathbf{x}_{M})$ provides a common center for each $p_{\mathbf{T}}(\mathbf{x}_{M})$ in the different target domains. This effectively anchors the residual functions and prevents overfitting to noisy samples.
\subsubsection{Scalable Distributional Regularization for Residual Learning}
\label{subsec:encoder}
Learning pairwise regularizers between each $p_{\mathbf{T}}(\mathbf{x}_{M})$ and the source input distribution $p_{\mathbf{S}}(\mathbf{x}_{M})$ is not a scalable solution. Instead we train a universal regularizer for each module $\mathcal{M}$ on the source $p_{\mathbf{S}}(\mathbf{x}_{\mathcal{M}})$, and apply this pre-trained regularizer when we fit the residual layers $\delta_{\mathcal{M}, \mathbf{T}}$ in each target domain. Our key intuition is to treat the regularizer for the inputs of each module $\mathcal{M}$ as a one-class decision-boundary~\cite{adv-novelty}, described by the dense regions in the source domain, i.e., $p_{\mathbf{S}}(\mathbf{x}_{\mathcal{M}})$. Unlike adversarial models that are trained with both the source and target distributions~\cite{adv-dist}, we propose a novel approach to learn distributional input regularizers for the shared modules with just the source domain inputs.
% $\mathbf{x}_{\mathcal{M}}^{\mathbf{T}} \sim p_{\mathbf{T}}(\mathbf{x}_{\mathcal{M}} + \delta^{\mathbf{T}}_{\mathcal{M}}(\mathbf{x}_{\mathcal{M}}))$.
%Maintaining cross-domain distributional consistency of the input representations $\mathbf{x}_{\mathcal{M}}$ to module $\mathcal{M}$ is involves fitting a  While generative adversarial approaches have proven effective at imitating distributions in the generated space~\cite{adv-dist, gan1}, adversarial models are trained jointly with both the source and target representations. In our application, 
For each shared module, the learned regularizer anticipates hard inputs across the target domains without accessing the actual samples. We introduce a variational encoder $\mathcal{E}_{\mathcal{M}}$ with RelU layers to map inputs $\mathbf{x}_{\mathcal{M}} \sim p_{\mathbf{S}}(\mathbf{x}_{\mathcal{M}})$ to a lower-dimensional reference distribution $\mathbf{N}(0, \mathbb{I})$~\cite{vae-tutorial}. Simultaneously, we add poisoning model $\mathcal{P}_{\mathcal{M}}$ to generate sample-adaptive noise $\mathcal{P}_{\mathcal{M}}(\mathbf{x}_{\mathcal{M}})$ to generate poisoned samples $\widetilde{\mathbf{x}}_{\mathcal{M}} = \mathbf{x}_{\mathcal{M}} + \mathcal{P}_{\mathcal{M}}(\mathbf{x}_{\mathcal{M}})$ with the source domain inputs $\mathbf{x}_{\mathcal{M}} \sim p_{\mathbf{S}}(\mathbf{x}_{\mathcal{M}})$. We define the encoder loss to train $\mathcal{E}_{\mathcal{M}}$ as follows:\vspace{1pt}
% so that $\widetilde{\mathbf{x}}_{\mathcal{M}} \sim \widetilde(p_{\mathbf{S}})$
\begin{equation}
\mathcal{L}_{\mathcal{E}_{\mathcal{M}}} = \infdiv{p(\mathcal{E}_{\mathcal{M}}(\mathbf{x}_{\mathcal{M}}))}{\mathbf{N}(0, \mathbb{I})} - \infdiv{p(\mathcal{E}_{\mathcal{M}}(\widetilde{\mathbf{x}}_{\mathcal{M}}))}{\mathbf{N}(0, \mathbb{I})}
\label{e_loss}
\end{equation}
where $\infdiv{p}{q}$ denotes the \textit{KL-Divergence} of distributions $p$ and $q$. The above loss enables $\mathcal{E}_{\mathcal{M}}$ to separate the true and poisoned samples across the $\mathbf{N}(0, \mathbb{I})$ hypersphere in its encoded space. Since $\mathcal{E}_{\mathcal{M}}(\mathbf{x}_{\mathcal{M}})$ involves a stochastic sampling step, gradients can be estimated with a \textit{reparametrization trick} using random samples to eliminate stochasticity in the loss $\mathcal{L}_{\mathcal{E}_{\mathcal{M}}}$\cite{vae-tutorial}. Conversely, the loss for our poisoning model $\mathcal{P}_{\mathcal{M}}$ is given by,
\vspace{-2pt}
\begin{equation}
\mathcal{L}_{\mathcal{P}_{\mathcal{M}}} = \infdiv{p(\mathcal{E}_{\mathcal{M}}(\widetilde{\mathbf{x}_{\mathcal{M}}}}{\mathbf{N}(0, \mathbb{I})}  - \log ||\mathcal{P}_{\mathcal{M}}(\mathbf{x}_{\mathcal{M}})||
\label{p_loss}
\end{equation}
Note the first term in \Cref{p_loss} attempts to confuse $\mathcal{E}_{\mathcal{M}}$ into encoding poisoned examples $\widetilde{\mathbf{x}_{\mathcal{M}}} = \mathbf{x}_{\mathcal{M}} + \mathcal{P}_{\mathcal{M}}(\mathbf{x}_{\mathcal{M}})$ in the reference distribution, while the second term prevents the degenerate solution $\mathcal{P}_{\mathcal{M}}(\mathbf{x}_{\mathcal{M}}) = 0$. \Cref{e_loss} and \Cref{p_loss} are alternatingly optimized, learning sharper decision boundaries as training proceeds. With the above alternating optimization, we pre-train the encoders $\mathcal{E}_{\mathcal{M}}$ for the three shared modules on the source domain $\mathbf{S}$. We now describe how we use these encoders to regularize the residual layers $\delta_{\mathcal{M}, \mathbf{T}}$ in each target domain $\mathbf{T}$.
\subsubsection{Distributionally-Regularized Target Loss}
For each target domain $\mathbf{T}$, we learn three residual layers for the module inputs $\mathbf{c}^{2}$, $\widetilde{\mathbf{e}_{u}}$ and $\widetilde{\mathbf{e}_{v}}$ for $\mathcal{M}_1, \mathcal{M}_3^{\mathcal{U}}, \mathcal{M}_3^{\mathcal{V}}$ respectively. The inputs to $\mathcal{M}_{4}$, $\widetilde{\mathbf{e}_{u}}^{n_{\mathcal{U}}}$,  $\widetilde{\mathbf{e}_{v}}^{n_{\mathcal{V}}}$  are not adapted. Thus, we learn three variational encoders in the source domain as described in \Cref{subsec:encoder}, $\mathcal{E}_{\mathbf{C}}, \mathcal{E}_{\mathcal{U}}$ and $\mathcal{E}_{\mathcal{V}}$ for $\mathbf{c}^{2}$, $\widetilde{\mathbf{e}_{u}}$ and $\widetilde{\mathbf{e}_{v}}$ respectively.
Consider target interactions $(u, \mathbf{c}, v) \in \mathcal{T}_{\mathbf{T}}$. In the absence of distributional regularization, the loss is identical to the first term in \Cref{negsample_loss}. However, we now apply regularizers to $\mathbf{c}^{2}$, $\widetilde{\mathbf{e}_{u}}, \widetilde{\mathbf{e}_{v}}$:
\begin{equation}
\begin{split}
\mathcal{L}^{reg}_{\mathcal{T}_{\mathbf{T}}} &= \mathcal{L}_{\mathcal{T}_{\mathbf{T}}} + \infdiv{p_{\mathbf{T}}(\mathcal{E}_{\mathcal{U}}({\widetilde{\mathbf{e}_{u}}}))}{\mathbf{N}(0, \mathbb{I})} + \\ & \infdiv{p_{\mathbf{T}}(\mathcal{E}_{\mathcal{V}}({\widetilde{\mathbf{e}_{v}}}))}{\mathbf{N}(0, \mathbb{I})} + \infdiv{p_{\mathbf{T}}(\mathcal{E}_{\mathbf{C}}({\mathbf{c}^{2}}))}{\mathbf{N}(0, \mathbb{I})}
\label{regloss}
\end{split}
\end{equation}
Again, the gradients can be estimated with the \textit{reparametrization trick} on the stochastic \textit{KL-divergence} terms\cite{vae-tutorial} as in \Cref{subsec:encoder}. The residual layers are then updated as in \Cref{subsec:Train} with $\mathcal{L}^{reg}_{\mathcal{T}_{\mathbf{T}}}$ replacing the first term in \Cref{negsample_loss}.

%% file: experiments.tex
\section{Experimental Results}
\label{sec:Experimental Results}
In this section, we present experimental analyses on diverse multi-domain recommendation datasets and show two key results. First, when we adapt modules trained on a rich source domain to the sparse target domains, we significantly reduce the computational costs and improve performance in comparison to learning directly on the sparse domains. Second, our model is comparable to \textit{state-of-the-art} baselines when trained on a single domain without transfer.
%\noindent
%\textbf{A: } Our recommender outperforms state-of-the-art baselines when trained in isolation, i.e., without meta-transfer.
%
%\noindent
%\textbf{B: } We observe performance and scalability gains under meta-transfer to the sparse targets vs. training them separately.
%
%We introduce datasets, baseline methods in \Cref{subsec:Datasets}, followed by the recommendation task and meta-transfer gains on sparse target domains in \Cref{subsec:Recommendation}, \Cref{subsec:Meta-Transfer-Exp}, qualitative analyses in \Cref{subsec:Qualitative} and scalability, robustness in~\Cref{subsec:Scale-Robust}. Finally, we discuss limitations and future directions in \Cref{subsec:Limitations}.
%We examine two critical questions: Q1---\textit{What is the effect of adversary weight $\lambda$ on interest space collapse and does this depend on the generator architecture?} and Q2---\textit{Is adversarial training robust to missing social or item history user data?} Finally, we analyze parameter sensitivity in \Cref{subsec:Parameter Sensitivity} and discuss limitations in~\Cref{subsec:Limitations}.
\begin{table}[b]
\centering
\vspace{-12pt}
\begin{tabular}{@{}p{0.120\linewidth}p{0.025\linewidth}p{0.240\linewidth}p{0.11\linewidth}p{0.10\linewidth}p{0.205\linewidth}@{}}
\toprule
\textbf{Dataset} &\textbf{}& \textbf{State} & \textbf{Users} & \textbf{Items} & \textbf{Interactions} \\ 
\midrule
 & \textbf{S} & \textbf{Bay-Area CA} & \textbf{1.20 m} & \textbf{8.90 k} & \textbf{25.0 m}  \\
\multirow{1}{*}{\textbf{FT-Data}} & $\mathbf{T}_1$ & Arkansas  & 0.40 m & 3.10 k & 5.20 m  \\
\multirow{1}{*}{$|\mathbf{C}| = 220$} & $\mathbf{T}_2$ & Kansas & 0.35 m & 2.90 k & 5.10 m  \\
& $\mathbf{T}_3$ & New-Mexico & 0.32 m & 2.80 k & 6.20 m  \\
& $\mathbf{T}_4$ & Iowa & 0.30 m & 3.00 k &  4.80 m  \\
\midrule
& \textbf{S} & \textbf{Pennsylvania} & \textbf{10.3 k} & \textbf{5.5 k} &\textbf{170 k}  \\
\multirow{1}{*}{\textbf{Yelp}}  & $\mathbf{T}_1$ & Alberta, Canada & 5.10 k & 3.5 k & 55.0 k  \\
\multirow{1}{*}{$|\mathbf{C}| = 120$} & $\mathbf{T}_2$ & Illinois  & 1.80 k& 1.05 k & 23.0 k  \\
& $\mathbf{T}_3$ & S.Carolina  & 0.60 k & 0.40 k & 6.20 k  \\
\midrule
  \multirow{1}{*}{\textbf{Google}}& \textbf{S} & \textbf{California} & \textbf{46 k} & \textbf{28 k} & \textbf{320 k}  \\
\multirow{1}{*}{\textbf{Local}} & $\mathbf{T}_1$ & Colorado & 10 k & 5.7 k & 51.0 k  \\
\multirow{1}{*}{$|\mathbf{C}| = 90$} & $\mathbf{T}_2$ & Michigan & 7.0 k & 4.0 k & 29.0 k  \\
& $\mathbf{T}_3$ & Ohio & 5.4 k & 3.2 k & 23.0 k  \\
\bottomrule
\end{tabular}
\label{data_stats}
\caption{\textit{Source} and \textit{Target} statistics for each of our datasets. Source states denoted S have more interactions and interaction density per user than target states denoted $\mathbf{T}_{i}$.}
\vspace{-14pt}
\end{table}
\subsection{Datasets and Baselines}
\label{subsec:Datasets}
We evaluate our recommendation model both with and without module transfer over the publicly available \textit{Yelp}\footnote{\url{https://www.yelp.com/dataset/challenge}} and \textit{Google Local Reviews}\footnote{\url{http://cseweb.ucsd.edu/~jmcauley/datasets.html}} datasets for benchmarking purposes. Reviews are split across U.S and Canadian states in these datasets. We treat each state as a separate recommendation domain for training and transfer purposes. There is no \textit{user or item overlap} across the states (recommendation domains) in any of our datasets. We repeat our experiments with a large-scale restaurant transaction dataset obtained from Visa (referred to as \textit{FT-Data}), also split across U.S. states. 
% They provide direct and inferred contextual features for each user rating (\textit{explicit feedback}) on businesses across FT-Data provides \textit{implicit feedback} to user (cardholder) preferences via transactions with businesses. Across all datasets, we observe significant performance gaps across states owing to variations in the interaction volume, density and quality. 
\begin{description}[leftmargin=0pt, labelindent=\parindent]
\item[\textbf{Google Local Reviews Dataset} (\textit{Explicit feedback})\footnote{\url{http://cseweb.ucsd.edu/~jmcauley/datasets.html}}\cite{mcauley1, mcauley2}:] Users rate businesses on a 0-5 scale with temporal, spatial, and textual context available for each review. We also infer additional context features - users' preferred locations on weekdays and weekends, spatial patterns and preferred product categories.
\item[\textbf{Yelp Challenge Dataset} (\textit{Explicit feedback}) \footnote{https://www.yelp.com/dataset/challenge}:] Users rate restaurants on a 0-5 scale, reviews include similar context features as the Google Local dataset. Further, \textit{user check-ins} and restaurant attributes (e.g., \textit{accepts-cards}) are available.
\item[\textbf{\textit{FT-Data}} (\textit{Implicit feedback}):] Contains the credit/debit card payments of users to restaurants in the U.S, with spatial, temporal, financial context features, and inferred transaction attributes. We leverage transaction histories also to infer user spending habits, restaurant popularity, peak hours, and tipping patterns.
\end{description}
In each dataset, we extract the same context features for every state with state-wise normalization, either with min-max normalization or quantile binning. We retain users and items with three or more reviews in the Google Local dataset and ten or more reviews in the Yelp dataset. In \textit{FT-Data}, we retain users and restaurants with over ten, twenty transactions, respectively, over three months. In each dataset, we choose a dense state with ample data as the source domain where all modules are trained, and multiple sparse states as target domains for module transfer from the source.
\vspace{-2pt}
\subsubsection{Source to Target Module Transfer}
\label{subsec:metabaselines}
We evaluate the performance gains obtained when we transfer or adapt modules $\mathcal{M}^1, \mathcal{M}^3$ and $\mathcal{M}^4$ from the source state to each target state, in comparison to training all four modules directly on the target. We also compare target domain gains with \textit{state-of-the-art} meta-learning baselines:
\begin{description}[leftmargin=0pt]
\item \textbf{LWA} \cite{nips17}: Learns a shared meta-model across all domains, with a user-specific linear component.
\item \textbf{NLBA} \cite{nips17}: Replaces LWA's linear component with a neural network with user-specific layer biases.
\item \textbf{$\mathbf{s^2}$-Meta} \cite{kdd19}: Develops a meta-learner to instantiate and train recommender models for each scenario. In our datasets, scenarios are the different states.
%\vspace{0.10in}
\item[\textbf{Direct Layer-Transfer (Our Variant)}:] Transfers source-trained meta-modules to the target-trained models as in \Cref{subsec:DLT}.
\item[\textbf{Anneal (Our Variant)}:] We apply simulated annealing to adapt the transferred meta-modules to the target as in \Cref{subsec:anneal}.
\item[\textbf{DRR - Distributionally Regularized Residuals}: ] \textbf{(Our Main Approach)} Adapts the inputs of each transferred module with separate residual layers in each target state. (\Cref{subsec:DRR}).
\end{description}
\vspace{-2pt}
\subsubsection{Single Domain Recommendation Performance}
We also evaluate the performance of our models independently without transfer on the source and target states in each dataset. We compare with the following \textit{state-of-the-art} recommendation baselines: 
%For each dataset, we train our recommender system on each state in isolation. When each model is trained and tested on its own state, the source-trained model significantly outperforms the target-only models. We compare our source model performance against , and demonstate the effectiveness of the proposed context transform model. We also experimentally validate that the learned transforms are generalizable and extensible to the target states. Our baselines are:
%% \vspace{-2pt}
%
\vspace{-2pt}
\begin{description}[leftmargin=0pt, labelindent=\parindent]
\item \textbf{NCF} \cite{ncf}: \textit{State-of-the-art} non context-aware model for comparisons and context validation.
\item \textbf{CAMF-C} \cite{camf}: Augments Matrix Factorization to incorporate a context-bias term for item latent factors. This version assumes a fixed bias for a given context feature for all items.
\item \textbf{CAMF} \cite{camf}: CAMF-C with separate context bias values for each item. We use this version for comparisons.
\item \textbf{MTF} \cite{ctf}: Obtains latent representations via decomposition of the User-Item-Context tensor. This model scales very poorly with the size of the context vector.
\item \textbf{NFM} \cite{nfm}:  Employs a bilinear interaction model applied to the context features of each interaction for representation. 
\item \textbf{AFM} \cite{afm}: Incorporates an attention mechanism to reweight the bilinear pooled factors in the NFM model. Scales poorly with the number of pooled contextual factors.
\item \textbf{AIN} \cite{cikm18context}: Reweights the interactions of user and item representations with each contextual factor via attention.
%\vspace{0.10in}
\item \textbf{MMT-Net (Our Main Approach)}: We refer to our model with all four modules as Multi-Linear Module Transfer Network (MMT-Net).
\item \textbf{FMT-Net (Our Variant)}: We replace $\mathcal{M}^1$s layers with feedforward \textit{RelU} layers to demonstrate the importance of multiplicative context invariants.
\item \textbf{MMT-Net Multimodal (Our Variant)}: \textbf{MMT-Net} with the information-gain terms described in \Cref{multimodal}. Only applied to \textit{FT-Data} due to lack of interactional features in other datasets.
\end{description}
\subsubsection{Experiment Setup}
We tune each baseline in parameter ranges centered at the author provided values for each dataset and set all embedding dimensions to $200$ for uniformity. We split each state in each dataset into training (80\%), validation (10\%), and test (10\%) sets for training, tuning, and testing purposes. 
For the \textit{implicit feedback} setting in \textit{FT-Data}, we adopt the standard negative-sample evaluation \cite{ncf} and draw one-hundred negatives per positive, equally split between item and context negatives similar to the training process in \Cref{subsec:Train}. We then evaluate the average \textbf{Hit-Rate@K} (\textbf{H@K}) metric for $K=1,5$ in \Cref{implicit_results}, indicating if the positive sample was ranked highly among the negative samples. For the \textit{explicit feedback} setting in the other two datasets, we follow the standard \textbf{RMSE} and \textbf{MAE} metrics in \Cref{explicit_results} \cite{camf, cikm18context} (no negative samples required). All models were implemented with \textit{Tensorflow} and tested on a \textit{Nvidia Tesla V100 GPU}. 
\begin{table}[hbtp]
 \centering
 \vspace{-6pt}
 \caption{Comparing aspects addressed by baseline recommendation models against our proposed \textbf{MMT-Net} approach}
 \vspace{-9pt}
 \begin{tabular}{@{}p{0.055\linewidth}p{0.165\linewidth}p{0.22\linewidth}p{0.08\linewidth}p{0.115\linewidth}p{0.18\linewidth}@{}}
  \toprule
     \textbf{} & \textbf{Bi-Linear Pooling} & \textbf{Multi-Linear Pooling} & \textbf{Low-Rank} & \textbf{Factor Weights} & \textbf{$\mathbf{\Theta}$(Context)}                                                                                                                           \\
  \midrule
\textbf{NFM} & Yes & No & No & No & Linear\\
\textbf{AFM} & Yes & No & No & Yes & Quadratic\\
\textbf{AIN} & No & No & Yes & Yes & Linear\\
\textbf{FMT} & No & No & Yes & Yes & Linear\\
\midrule
\textbf{MMT} & Yes & Yes & Yes & Yes & Linear\\
  \bottomrule
 \end{tabular}
 \label{methods}
 \vspace{-12pt}
\end{table}
\subsection{Module Transfer to Sparse Target States}
\label{subsec:Meta-Transfer-Exp}
\begin{table}[b]
\vspace{-10pt}
\centering
\begin{tabular}{@{}p{0.120\linewidth}p{0.03\linewidth}p{0.085\linewidth}p{0.1\linewidth}p{0.080\linewidth}p{0.080\linewidth}p{0.085\linewidth}p{0.140\linewidth}@{}}
\toprule
\textbf{Dataset} & \textbf{} & \textbf{Direct} & \textbf{Anneal} & \textbf{DRR} & \textbf{LWA} & \textbf{NLBA}  & $\mathbf{s^2}$\textbf{-Meta}   \\ 
\small\textbf{} & \small\textbf{} & \small\textbf{\%H@1} & \small\textbf{\%H@1} & \small\textbf{\%H@1} & \small\textbf{\%H@1} & \small\textbf{\%H@1} & \small\textbf{\%H@1} \\
\midrule
\multirow{4}{*}{\textbf{FT-Data}} & $T_1$ & 2\% & \textbf{19\%}  & 18\%  & 6\%  & x & x \\
& $T_2$ & 0\%  & \textbf{16\%}  & \textbf{16\%}  & 8\%  & x & x \\
& $T_3$ & 3\%  & \textbf{18\%}  & \textbf{18\%}  & 6\%   & x & x \\
& $T_4$ & -1\%  & \textbf{14\%} &  12\%  & 11\%  & x & x \\
\bottomrule
\end{tabular}
\vspace{2pt}
\caption{Percentage improvements (\textbf{\% Hit-Rate@1}) on \textit{FT-Data} target states with module transfer approaches and meta-learning baselines against training all modules on the target state directly as in \Cref{implicit_results}.}
\label{implicit_meta_results}
\vspace{-21pt}
\end{table}
We evaluate module transfer methods by the percentage improvements in the \textbf{Hit-Rate@1} for the implicit feedback setting in \textit{FT-Data} (\Cref{implicit_meta_results}), or the drop in \textbf{RMSE} (\Cref{explicit_meta_results}) for the explicit feedback datasets when we transfer the $\mathcal{M}^1, \mathcal{M}^3$ and $\mathcal{M}^4$ modules from the source state rather than training all four modules from scratch on that target domain. Similarly, meta-learning baselines were evaluated by comparing their joint meta-model performance on the target state against our model trained only on that state. The performance numbers for training our model on each target state without transfer are recorded in \Cref{explicit_results}, \Cref{implicit_results}.
We could not scale the NLBA, LWA and \textbf{$\mathbf{s^2}$-Meta} approaches to \textit{FT-Data} owing to the costs of training the meta-models on all users combined across the source and multiple target domains. In \Cref{explicit_meta_results}, we demonstrate the percentage reduction in RMSE with module transfer for Google Local, Yelp, and in \Cref{implicit_meta_results}, we demonstrate significant improvements in the hit-rates for \textit{FT-Data}. We start with an analysis of the training process for module transfer with simulated annealing and DRR adaptation.
%\begin{figure}
%\includegraphics[width=0.89\linewidth]{images/plots/scaling-image.pdf}
%\caption{.}
%\label{scaling-image}
%\end{figure}
%\begin{figure}
%\includegraphics[width=0.89\linewidth]{images/plots/Google-Local-MI.pdf}
%\label{Google-Local-MI}
%\caption{The Train-RMSE values of MMT-Net when trained on just the target (Google-Local MI), when annealed and learned with residuals after 2 epochs of pre-training.}
%\end{figure}
%
%\begin{figure}
%\includegraphics[width=0.89\linewidth]{images/plots/Google-Local-OH.pdf}
%\label{Google-Local-OH}
%\caption{The Train-RMSE values of MMT-Net when trained on just the target (Google-Local OH), when annealed and learned with residuals after 2 epochs of pre-training.}
%\end{figure}
\begin{table}[t]
\centering
%\begin{tabular}{@{}p{0.102\linewidth}p{0.05\linewidth}p{0.039\linewidth}p{0.039\linewidth}p{0.039\linewidth}p{0.039\linewidth}p{0.039\linewidth}p{0.039\linewidth}p{0.039\linewidth}p{0.039\linewidth}p{0.039\linewidth}p{0.039\linewidth}p{0.039\linewidth}p{0.039\linewidth}@{}}
%\toprule
%\textbf{Dataset} & \textbf{Target} & \multicolumn{2}{c}{\textbf{Direct}} & \multicolumn{2}{c}{\textbf{Anneal}} & \multicolumn{2}{c}{\textbf{DRR}} & \multicolumn{2}{c}{\textbf{LWA}}  & \multicolumn{2}{c}{\textbf{NLBA}}  & \multicolumn{2}{c}{$\mathbf{s^2}$\textbf{-Meta}}   \\ 
%
%\textit{} & \textit{} & \textit{RMS} & \textit{MAE} & \textit{RMS} & \textit{MAE} & \textit{RMS} & \textit{MAE} & \textit{RMS} & \textit{MAE} &\textit{RMS} & \textit{MAE} & \textit{RMS} & \textit{MAE} \\
\caption{Percentage RMSE improvements on the Yelp and Google Local target states with module transfer approaches and meta-learning baselines against training all modules on the target state directly as in \Cref{explicit_results}.}
\vspace{-10pt}
\begin{tabular}{@{}p{0.12\linewidth}p{0.03\linewidth}p{0.085\linewidth}p{0.1\linewidth}p{0.080\linewidth}p{0.080\linewidth}p{0.085\linewidth}p{0.140\linewidth}@{}}
\toprule
\textbf{Dataset} & \textbf{} & \textbf{Direct} & \textbf{Anneal} & \textbf{DRR} & \textbf{LWA} & \textbf{NLBA}  & $\mathbf{s^2}$\textbf{-Meta}   \\ 
\small\textbf{} & \small\textbf{} & \small\textbf{\%RMSE} & \small\textbf{\%RMSE} & \small\textbf{\%RMSE} & \small\textbf{\%RMSE} & \small\textbf{\%RMSE} & \small\textbf{\%RMSE} \\
\midrule
\multirow{3}{*}{\textbf{Yelp}} & $T_1$ & -2.2\% &\textbf{ 7.7\%}  &7.2\% & 2.6\% & 4.1\% & 3.7\%  \\
& $T_2$ & -2.6\% & \textbf{9.0\%}  & 7.9\%& 1.8\% & 3.6\%  & 3.1\%  \\
& $T_3$ & 0.8\%  & \textbf{8.5\%}  & 8.1\%  & 0.3\%  & 5.3\%  & 1.8\%  \\
\midrule
\multirow{1}{*}{\parbox{0.12\linewidth}{\textbf{Google}}} & $T_1$ & -1.2\%  & \textbf{11.2\%}  & 11.0\%  & 3.3\% & 4.3\%  & 3.1\%  \\
\textbf{Local} & $T_2$ & -1.7\%  & \textbf{12.1\%}  & 10.9\%  & 4.6\%  & 4.9\% & 2.8\%\\
 & $T_3$ & -2.0\%  & \textbf{9.6\%}  & 8.8\%  & 2.4\%  & 6.3\% & 3.9\%  \\
\bottomrule
\end{tabular}
\label{explicit_meta_results}
\vspace{-22pt}
\end{table}

\textbf{Transfer Details: } On each target state in each dataset, all four modules of our MMT-Net model are pretrained over two gradient epochs on the target samples. The layers in modules $\mathcal{M}^1, \mathcal{M}^3$ and $\mathcal{M}^4$ are then replaced with those trained on the source state, while retaining module $\mathcal{M}^2$ without any changes (in our experiments $\mathcal{M}^2$ just contains user and item embeddings, but could also include neural layers if required). This is then followed by either simulated annealing or DRR adaptation of the transferred modules. We analyze the training loss curves in \Cref{subsec:convergence} to better understand the fast adaptation of the transferred modules. 

\textbf{Invariant Quality: }A surprising result was the similar performance of \textit{direct layer-transfer} with no adaptation to training all modules on the target state from scratch (\Cref{explicit_meta_results}). The transferred source state modules were directly applicable to the target state embeddings. This helps us validate the generalizability of context-based modules across independently trained state models even with no user or item overlap.

\textbf{Computational Gains: }We also plot the total training times including pretraining for DRR and annealing against the total number of target state interactions in \Cref{scaling-image}. On the target states, module transfer is ~3x faster then training all the modules from scratch. On the whole, there is a significant reduction in the overall training time and computational effort in the \textit{one-to-many} setting. Simulated annealing and DRR adaptation converge in fewer epochs when applied to the pre-trained target model, and outperform the target-trained model by significant margins (\Cref{explicit_meta_results}). These computational gains potentially enable a finer target domain granularity (e.g., adapt to towns or counties rather than states).
\begin{table*}[t]
\centering
\caption{We evaluate recommendation performance on each state (no transfer) with RMSE, MAE metrics for \textit{explicit feedback} against the ground-truth ratings. Metrics were averaged over five runs, $*$ indicates statistical significance (paired \textit{t-test}, p=0.05). On average, models incorporating both pooling and reweighting in \Cref{methods} exhibit significant relative gains (i.e., \textbf{AFM}, \textbf{MMT}).}
\vspace{-11pt}
\begin{tabular}{@{}p{0.06\linewidth}p{0.04\linewidth}p{0.0355\linewidth}p{0.0355\linewidth}p{0.0355\linewidth}p{0.0355\linewidth}p{0.0355\linewidth}p{0.0355\linewidth}p{0.0355\linewidth}p{0.0355\linewidth}p{0.0355\linewidth}p{0.0355\linewidth}p{0.0355\linewidth}p{0.0355\linewidth}p{0.0355\linewidth}p{0.0355\linewidth}p{0.0355\linewidth}p{0.0355\linewidth}@{}}
\toprule
\textbf{Dataset} & \textbf{State} & \multicolumn{2}{c}{\textbf{CAMF}~\cite{camf}} & \multicolumn{2}{c}{\textbf{MTF}~\cite{ctf}} &  \multicolumn{2}{c}{\textbf{NCF}~\cite{ncf}} & \multicolumn{2}{c}{\textbf{NFM}~\cite{nfm}}  & \multicolumn{2}{c}{\textbf{AFM}~\cite{afm}}  & \multicolumn{2}{c}{\textbf{AIN}~\cite{cikm18context}}  & \multicolumn{2}{c}{\textbf{FMT-Net}}  & \multicolumn{2}{c}{\textbf{MMT-Net}} \\ 
\small\textbf{} & \small\textbf{} & \small\textbf{RMS} & \small\textbf{MAE} & \small\textbf{RMS} & \small\textbf{MAE} & \small\textbf{RMS} & \small\textbf{MAE} & \small\textbf{RMS} & \small\textbf{MAE} & \small\textbf{RMS} & \small\textbf{MAE} &\small\textbf{RMS} & \small\textbf{MAE} & \small\textbf{RMS} & \small\textbf{MAE} & \small\textbf{RMS} & \small\textbf{MAE} \\
\midrule
\multirow{4}{*}{\textbf{Yelp}} & \textbf{S} & 1.21 & 0.94 & 1.13 & 0.87 & 1.18 & 1.04 & 1.02 & 0.83 & 0.96 & 0.78 & 0.98 & 0.75 & 1.02 & 0.76 & \textbf{0.94} & \textbf{0.73}\\
\midrule
& $T_1$ & 1.56 & 1.20 & 1.41 & 1.12 & 1.39 & 0.99 &  1.29 & 1.01 & 1.27 & 0.94 & 1.36 & 0.91 & 1.34 &  0.95& \textbf{1.24}* & \textbf{0.88}*\\
& $T_2$ & 1.33 & 1.04 & 1.36 & 0.98 & 1.26 & 1.02 &  1.19 & 1.05 & 1.16 & \textbf{0.90} & 1.17 & 0.95 & 1.15 & 0.98 & \textbf{1.13}* & 0.91\\
& $T_3$ & 1.49 & 1.13 & 1.50 & 1.08 & 1.35 & 1.08 &  1.31 & 0.96 & \textbf{1.20}* & 0.93 & 1.25 & 0.98 & 1.29 & 1.02 & \textbf{1.20}* & \textbf{0.89}*\\
\midrule
 & \textbf{S} & 1.36 & 1.01 & 1.21 & 0.90 & 1.04 & 0.89 & 0.80 & 0.73 & \textbf{0.77} & 0.63 & 0.85 & 0.64 & 0.91 & 0.68 & \textbf{0.77} & \textbf{0.64} \\
\midrule
\multirow{1}{*}{\textbf{Google}} & $T_1$ & 1.49 & 1.20 & 1.38 & 1.14& 1.27 & 1.05 & 1.10 & 0.99 & 0.94 & 0.85 & 1.22 & 0.90 & 1.31 & 0.96 & \textbf{0.89} & \textbf{0.76}* \\
\multirow{1}{*}{\textbf{Local}} & $T_2$ & 1.37 & 1.16 & 1.31 & 1.20& 1.36 & 1.17 & 1.21 & 1.05 &  \textbf{1.14}* & 0.98 & 1.19 & 1.01 & 1.28 & 1.07 & 1.16 & \textbf{0.93}*\\
& $T_3$ & 1.39 & 1.23 & 1.20 & 1.07& 1.19 & 0.98 & 1.13 & 0.92 & 1.09 & 0.91 & 1.08 & 0.94 & 1.14 & 0.98 & \textbf{1.02}* & \textbf{0.85}* \\
\bottomrule
\end{tabular}
\label{explicit_results}
\vspace{-4pt}
\end{table*}
\begin{table*}[b]
\centering
\begin{tabular}{@{}p{0.06\linewidth}p{0.04\linewidth}|p{0.02\linewidth}p{0.02\linewidth}p{0.0375\linewidth}p{0.0375\linewidth}p{0.0375\linewidth}p{0.0375\linewidth}p{0.02\linewidth}p{0.0375\linewidth}p{0.0375\linewidth}p{0.0375\linewidth}p{0.0375\linewidth}p{0.0375\linewidth}p{0.0375\linewidth}p{0.0375\linewidth}p{0.0375\linewidth}@{}}
\toprule
\textbf{Dataset} & \textbf{State} & \multicolumn{1}{c}{\textbf{CAMF}} & \multicolumn{1}{c}{\textbf{MTF}} & \multicolumn{2}{c}{\textbf{NCF}~\cite{ncf}} & \multicolumn{2}{c}{\textbf{NFM}~\cite{nfm}}  & \multicolumn{1}{c}{\textbf{AFM}}  & \multicolumn{2}{c}{\textbf{AIN}~\cite{cikm18context}}  & \multicolumn{2}{c}{\textbf{FMT-Net}}  & \multicolumn{2}{c}{\textbf{MMT-Net}} & \multicolumn{2}{c}{\textbf{MMT-m}} \\ 
\textbf{} & \textbf{} & \cite{camf} &\cite{ctf} & \small\textbf{H@1} & \small\textbf{H@5} & \small\textbf{H@1} & \small\textbf{H@5} & \cite{afm} & \small\textbf{H@1} & \small\textbf{H@5} & \small\textbf{H@1} & \small\textbf{H@5} & \small\textbf{H@1} & \small\textbf{H@5} & \small\textbf{H@1} & \small\textbf{H@5}\\
\midrule
\multirow{5}{*}{\textbf{FT-Data}} & \textbf{S} & x & x & 0.42 & 0.77  & 0.52 & 0.91 & x & 0.44 & 0.89 & 0.37 & 0.76 & \textbf{0.56}* & \textbf{0.94} & \textbf{0.56}* & 0.93\\
\midrule
& $T_1$ & x & x & 0.36 & 0.71 & 0.41 & 0.83 & x & 0.34 & 0.76 & 0.32 & 0.75 &  0.45 & 0.84 & \textbf{0.47}* & \textbf{0.86}* \\
& $T_2$  & x & x & 0.25 & 0.64 & 0.30 & 0.77 & x & 0.30 & 0.72 & 0.26 & 0.72 &  \textbf{0.34}* & \textbf{0.79} & \textbf{0.34}* & 0.77 \\
& $T_3$  & x & x & 0.26 & 0.70 & 0.31 & 0.78 & x & 0.29 & 0.74 & 0.28 & 0.74 &  0.33 & \textbf{0.82}* & \textbf{0.34} & 0.80\\
& $T_4$  & x & x & 0.29 & 0.72 & 0.32 & 0.74 & x & 0.32 & 0.78 & 0.21 & 0.69 & 0.37 & 0.80 & \textbf{0.38} & \textbf{0.83}*\\
\bottomrule
\end{tabular}
\caption{We evaluate recommendation performance on each state (no transfer) with the \textit{H@1, 5} metrics for \textit{implicit feedback} in \textit{FT-Data}. Metrics were averaged over five runs, * indicates statistical significance (paired \textit{t-test}, p=0.05). On average, feature-pooling methods AFM, NFM and MMT outperform additive models AIN, FMT. x indicates timed-out or memory limit exceeded.}
\label{implicit_results}
\vspace{-20pt}
\end{table*}
\subsection{Single Domain Recommendation}
\label{subsec:Recommendation}
\subsubsection{Comparative Analysis} 
We draw attention to the most relevant features of the baselines and our variants in \Cref{methods}. We highlight our key observations from the experimental results obtained with the baseline recommenders and our FMT-Net and MMT-Net variants (\Cref{implicit_results}, \Cref{explicit_results}). Note that methods with some form of context pooling significantly outperform those without pooled factors, indicating the importance of multi-linear model expressivity to handle interaction context.

We note that AFM performs competitively owing to its ability to reweight terms, similar to our approach (\Cref{explicit_results}), but fails to scale to the larger \textit{FT-Data}. NFM is linear with context size owing to a simple algebraic re-organization, and thus scales to \textit{FT-Data}, however losing the ability to reweight pairwise context product terms \cite{nfm}. Also note the differences between our FMT and MMT variants, demonstrating the importance of the pooled multi-linear formulation for the contextual invariants. These performance differences are more pronounced in the implicit feedback setting (\Cref{implicit_results}). This can be attributed to the greater relevance of transaction context (e.g., transactions provide accurate temporal features while review time is a proxy to the actual visit) and more context features in \textit{FT-Data} vs. Google Local and Yelp (220 vs. 90,120 respectively), magnifying the importance of feature pooling for \textit{FT-Data}.

The lack of pooled feature expressivity in the FMT-Net model impacts the training process as seen in \Cref{pooling-image}, demonstrating the importance of context intersection. The NFM and MMT models converge faster to a smaller Train-RMSE in \Cref{pooling-image} and outperform FMT on the test data (\Cref{implicit_results}, \Cref{explicit_results}). We also observe models incorporating pooled factors to outperform the inherently linear attention-based AIN model, although the performance gap is less pronounced in the smaller review datasets (\Cref{explicit_results}).  We now qualitatively analyze our results to interpret module adaptation.
\subsection{Qualitative Analysis}
\label{subsec:Qualitative}
We analyze our approach from the shared module training and convergence perspective for the different adaptation methods. We observe consistent trends across the direct layer-transfer, annealing, and DRR adaptation approaches. 
\subsubsection{Training without Context-Bias} 
\begin{figure}[t]
\vspace{-10pt}
%\hfill
\includegraphics[clip, trim=0cm 0cm 0cm 1cm, width=0.78\linewidth]{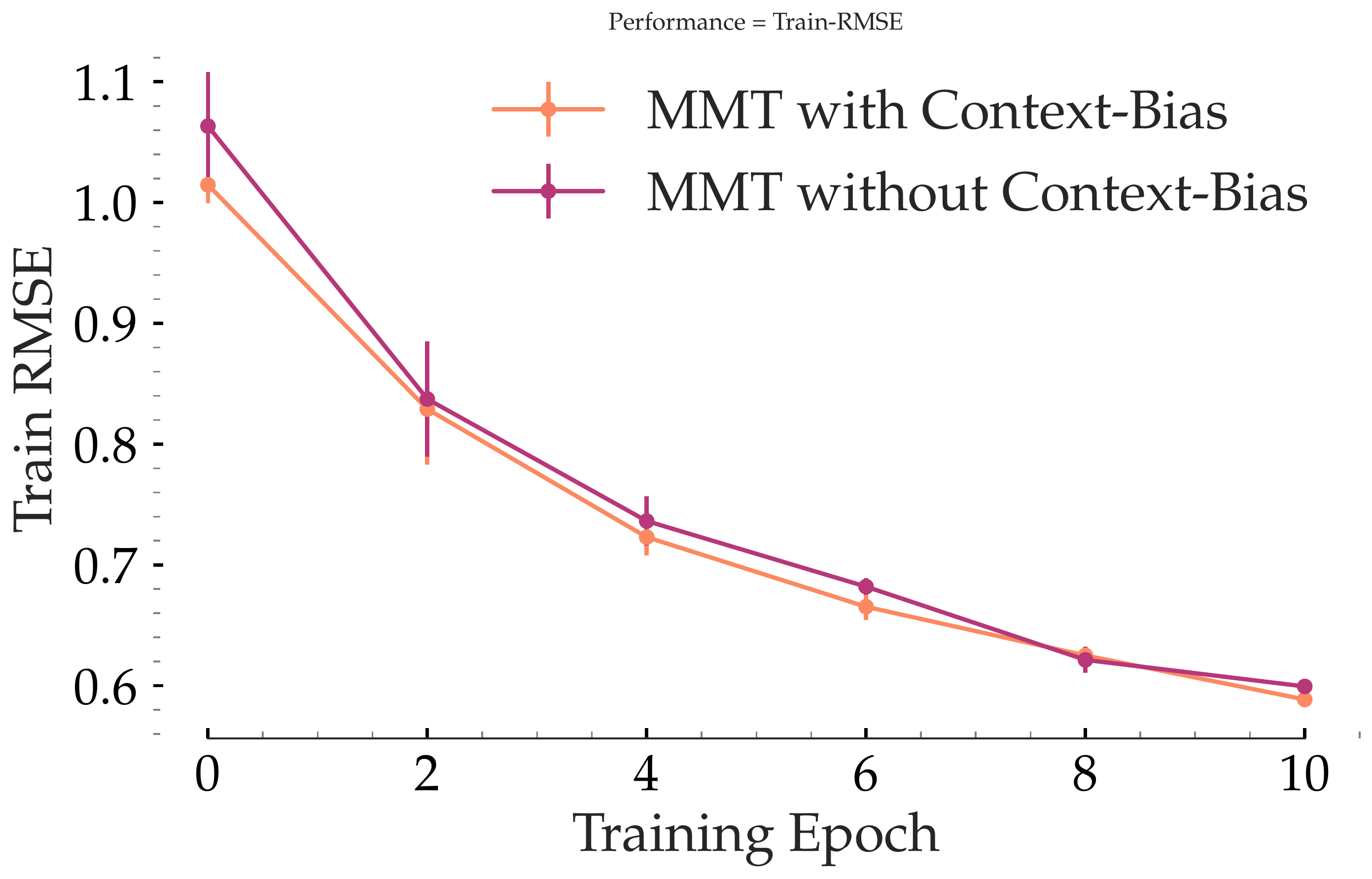}
%\hfill
%\subfigure[MMT-Net convergence with and without context-bias]{\includegraphics[clip, trim=0cm 0cm 0cm 1cm, width=0.49\linewidth]{images/plots/Train-MMT-context.pdf}}
%\hfill
%%\caption{Overall Performance and Percentage Gains of Asr-VAE (by R@50), measured across user quantiles. Clearly, users in the third or fourth quartiles of either axis receive better overall recommendations. However, performance gains are mostly observed for users in the first and second item count quartiles (sparse histories)}%
\vspace{-12pt}
\caption{MMT-Net trained with \& without context-bias (\Cref{eq:context_bias}) on the Google Local source exhibits similar Train-RMSE, but registers $>10\%$ drop in test performance.}
\vspace{-14pt}
\label{train-convergence-context}
\end{figure}

To understand the importance of decorrelating training samples in the training process, we repeat the performance analysis on our MMT-Net model with and without the adaptive context-bias term in the training objective in \Cref{subsec:Train}. We observe a 15\% performance drop across the Yelp and Google Local datasets, although this does not reflect in the Train-RMSE convergence (\Cref{train-convergence-context}) of the two variations. In the absence of context-bias, the model overfits uninformative transactions to the user and item bias terms ($s_{u}$, $s_{v}$) in \Cref{explicit_rating}, \Cref{explicit_loss}  and thus achieves comparable Train-RMSE values. However, the overfit user and item terms are not generalizable, resulting in the observed drop in test performance.
\subsubsection{Model Training and Convergence Analysis}
\label{subsec:convergence}
\begin{figure}[b]
\vspace{-4pt}
%\hfill
\includegraphics[clip, trim=0cm 0cm 0cm 0.7cm, width=0.78\linewidth]{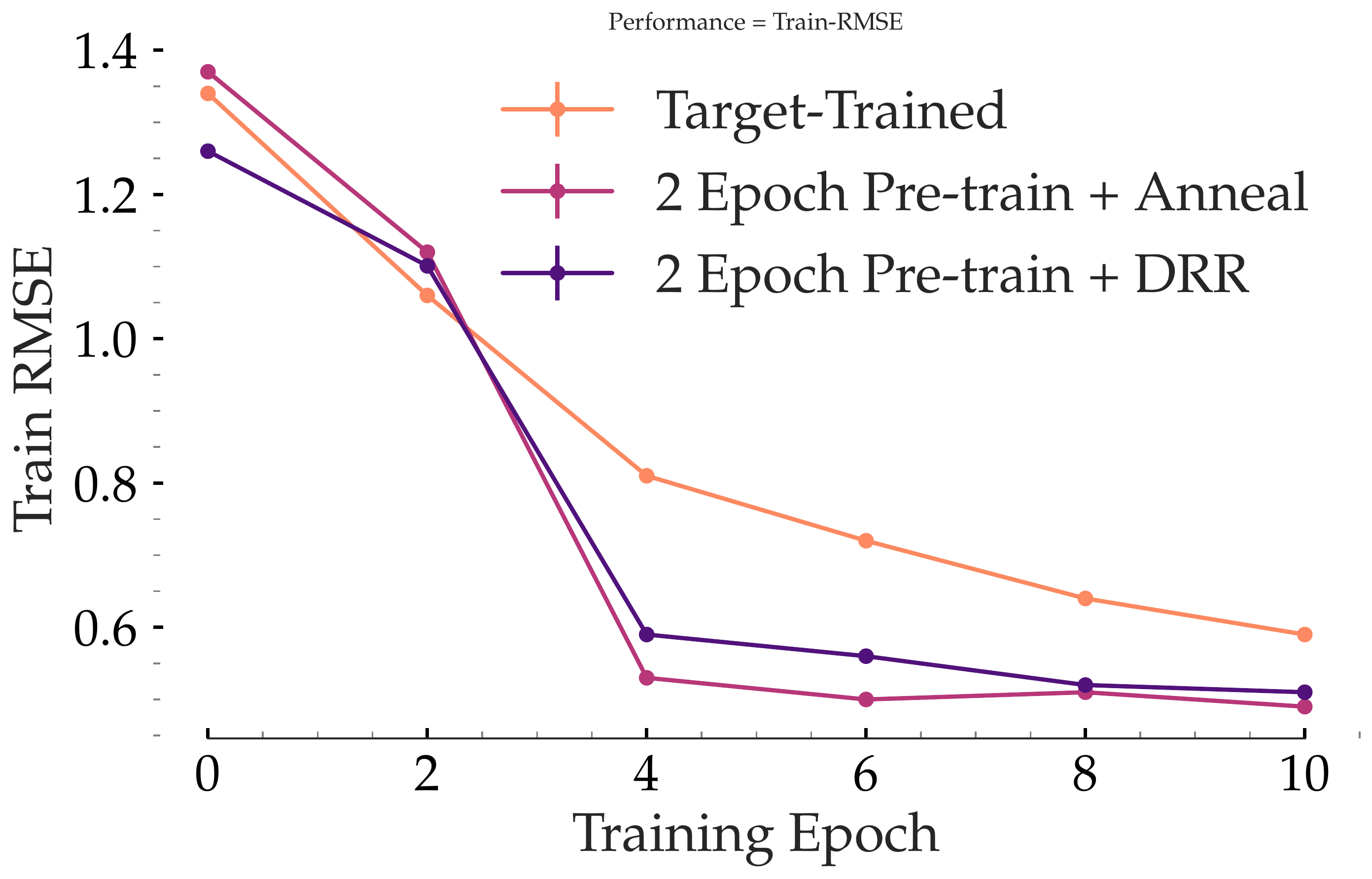}
%\hfill
%\subfigure[MMT-Net convergence with and without context-bias]{\includegraphics[clip, trim=0cm 0cm 0cm 1cm, width=0.49\linewidth]{images/plots/Train-MMT-context.pdf}}
%\hfill
%%\caption{Overall Performance and Percentage Gains of Asr-VAE (by R@50), measured across user quantiles. Clearly, users in the third or fourth quartiles of either axis receive better overall recommendations. However, performance gains are mostly observed for users in the first and second item count quartiles (sparse histories)}%
\vspace{-12pt}
\caption{MMT-Net convergence under target-training vs. Annealing/DRR adaptation after 2 epochs of pretraining on the Google Local Colorado target}
\vspace{-12pt}
\label{fig:convergence}
\end{figure}

\begin{figure}[b]
\vspace{-2pt}
%\hfill
\includegraphics[clip, trim=0cm 0cm 0cm 1cm, width=0.78\linewidth]{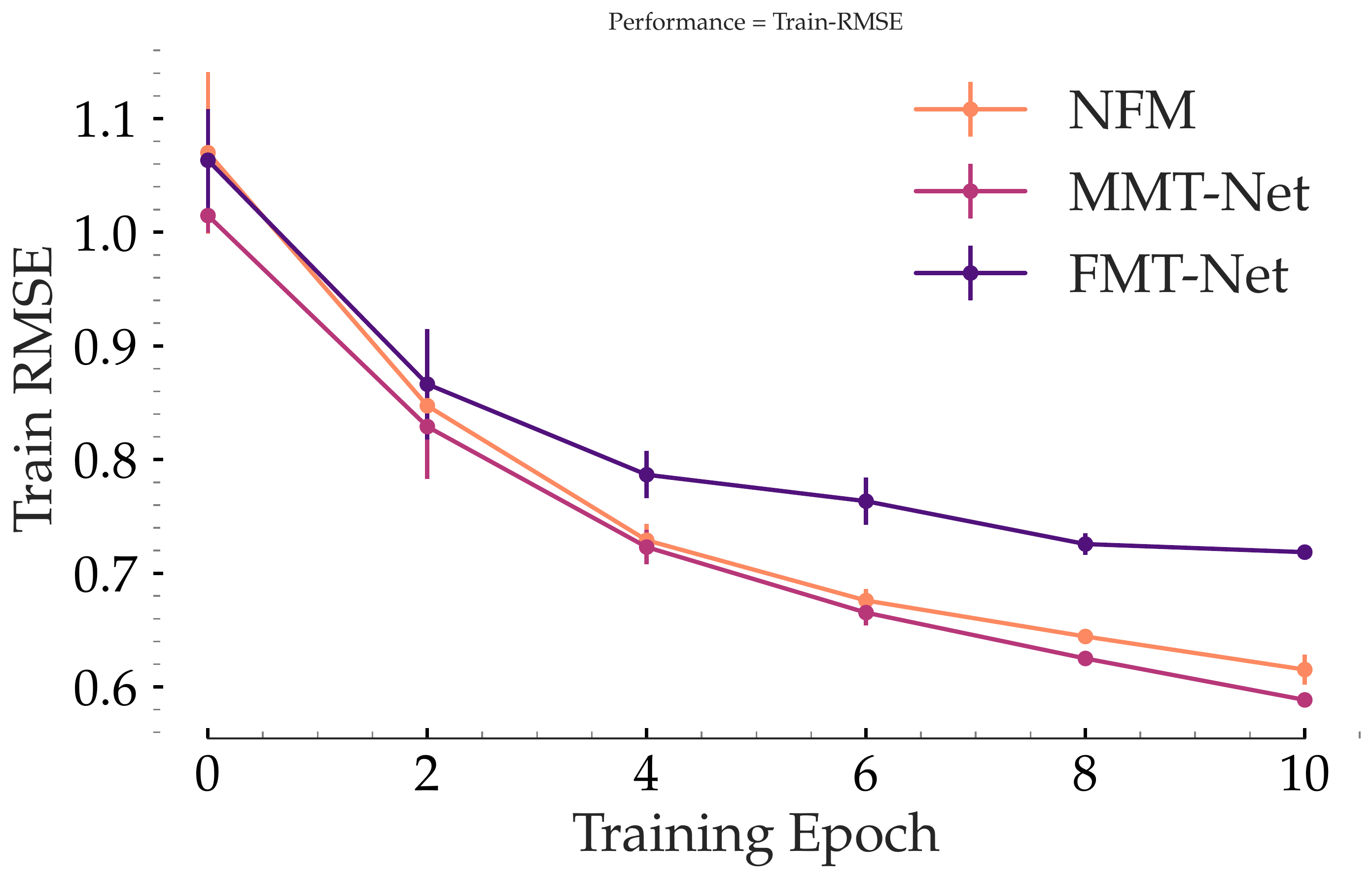}
%\hfill
%\subfigure[MMT-Net convergence with and without context-bias]{\includegraphics[clip, trim=0cm 0cm 0cm 1cm, width=0.49\linewidth]{images/plots/Train-MMT-context.pdf}}
%\hfill
%%\caption{Overall Performance and Percentage Gains of Asr-VAE (by R@50), measured across user quantiles. Clearly, users in the third or fourth quartiles of either axis receive better overall recommendations. However, performance gains are mostly observed for users in the first and second item count quartiles (sparse histories)}%
\vspace{-12pt}
\caption{MMT-Net convergence compared to NFM and FMT-Net on the Google Local Colorado target}
%\vspace{-25pt}
\label{pooling-image}
\end{figure}

\begin{figure}[t]
\vspace{-10pt}
%\hfill
\includegraphics[clip, trim=0cm 0cm 0cm 1cm, width=0.78\linewidth]{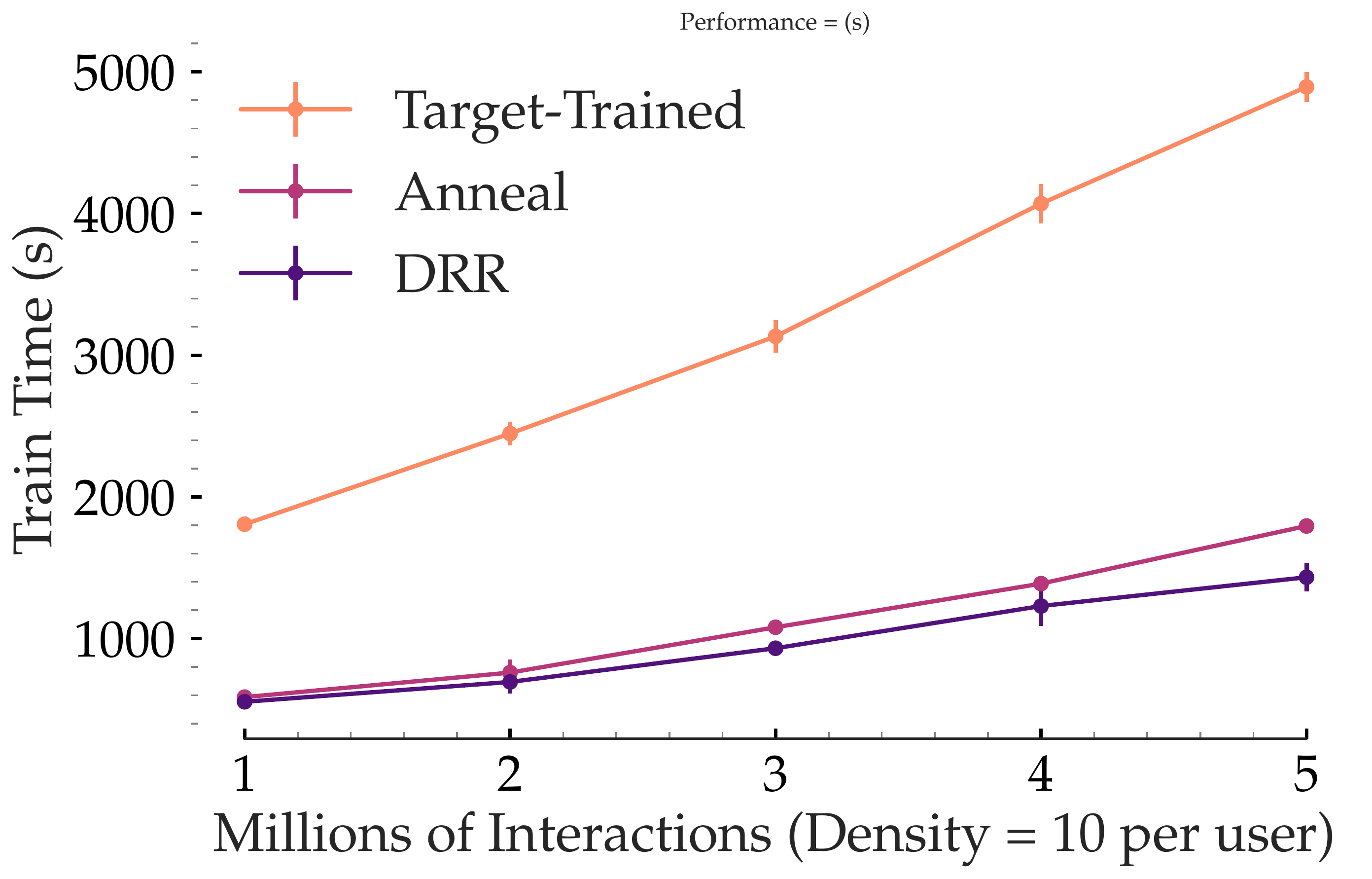}
%\hfill
%\subfigure[MMT-Net convergence with and without context-bias]{\includegraphics[clip, trim=0cm 0cm 0cm 1cm, width=0.49\linewidth]{images/plots/Train-MMT-context.pdf}}
%\hfill
%%\caption{Overall Performance and Percentage Gains of Asr-VAE (by R@50), measured across user quantiles. Clearly, users in the third or fourth quartiles of either axis receive better overall recommendations. However, performance gains are mostly observed for users in the first and second item count quartiles (sparse histories)}%
\vspace{-12pt}
\caption{MMT-Net training duration with and without module transfer vs. target domain interaction volume}
\vspace{-14pt}
\label{scaling-image}
\end{figure}

%We highlight a few consistent observations across datasets and target domains:
%\begin{itemize}
We compare the Train-RMSE convergence for the MMT-Net model fitted from scratch to the Google Local target state, Colorado ($\mathbf{T}_1$) vs. the training curve under DRR and annealing adaptation with two pretraining epochs on the target state in \Cref{fig:convergence}. Clearly, the target-trained model takes significantly longer to converge to a stable Train-RMSE in comparison to the Anneal and DRR adaptation. Although the final Train-RMSE is comparable (\Cref{scaling-image}), there is a significant performance difference between the two approaches on the test dataset, as observed in \Cref{explicit_meta_results}. Training loss convergence alone is not indicative of the final model performance; the target-only training method observes lower Train-RMSE by overfitting to the sparse data. We also compare the Train-RMSE convergence for target-trained models with and without pooled context factors (MMT-Net, NFM vs. FMT-Net) in \Cref{pooling-image}. We observe the NFM, MMT-Net models to converge faster to a better optimization minima than FMT-Net. This also reflects in their test performance in \Cref{implicit_results}.
\subsection{Scalability and Robustness Analysis}
\label{subsec:Scale-Robust}
We demonstrate the scalability of meta-transfer with the number of transactions in the target domain in \Cref{scaling-image} against training separate models. Our previous observations in \Cref{subsec:Meta-Transfer-Exp} validate the ability of our approach to scale deeper architectures to a large number of target domains while also enabling a finer resolution for the selection of target domains. Towards tackling incomplete data, we also evaluated the robustness of the shared context layers by randomly dropping up to 20\% of the context features in each interaction at train and test time for both, the source and target states in \Cref{table-context-drop}. 
%\begin{figure}
%\includegraphics[width=0.69\linewidth]{images/plots/scalability-numbers.pdf}
%
%\caption{Training time for the Target-Only, Anneal and DRR approaches against Millions of Interactions (with user interaction density of 10)}
%\label{scaling-image}
%\end{figure}
\begin{table}[hbtp]
 \centering
 \vspace{-8pt}
 \caption{MMT-Net performance degradation was measured by the decrease in \textbf{H@1} or increase in \textbf{RMSE}, averaged over target states with random context feature dropout}
 \vspace{-11pt}
 \begin{tabular}{@{}p{0.25\linewidth}p{0.17\linewidth}p{0.16\linewidth}p{0.16\linewidth}p{0.16\linewidth}@{}}
  \toprule
     \textbf{Context Drop} & \textbf{5\%} & \textbf{10\%} & \textbf{15\%} & \textbf{20\%} \\
  \midrule
\textbf{FT-Data} & 1.1\% & 2.6\% & 4.1\% & 6.0\% \\
\textbf{Google Local} & 3.9\% & 4.2\% & 7.0\% & 8.8\% \\
\textbf{Yelp} & 1.8\% & 3.2\% & 5.4\% & 7.3\% \\
  \bottomrule
 \end{tabular}
 \label{table-context-drop}
  \vspace{-18pt}
 \end{table}
% (\Cref{}). These results support the generalizability and scalability of our approach to recommendation applications.
\subsection{Limitations and Discussion}
\label{subsec:Limitations}
We identify a few fundamental limitations of our model. While our approach presents a scalable and effective solution to bridge the weaknesses of gradient-based meta learning and co-clustering via user or item overlaps, contextual invariants do not extend to cold-start users or items. Second, our model does not trivially extend to the case where a significant number of users or items are shared across recommendation domains. We separate the embeddings and $\textit{learn-to-learn}$ aspect which improves modularity, but prevents direct reuse of representations across domains, since only the transformation layers are shared. Depending on the application, context features could potentially be filtered to enhance social inference and prevent loss of diversity in the generated recommendations.

%% file: fp0047.bbl

\begin{thebibliography}{50}

%%% the Web address.

\ifx \showCODEN    \undefined \def \showCODEN     #1{\unskip}     \fi
\ifx \showDOI      \undefined \def \showDOI       #1{#1}\fi
\ifx \showISBNx    \undefined \def \showISBNx     #1{\unskip}     \fi
\ifx \showISBNxiii \undefined \def \showISBNxiii  #1{\unskip}     \fi
\ifx \showISSN     \undefined \def \showISSN      #1{\unskip}     \fi
\ifx \showLCCN     \undefined \def \showLCCN      #1{\unskip}     \fi
\ifx \shownote     \undefined \def \shownote      #1{#1}          \fi
\ifx \showarticletitle \undefined \def \showarticletitle #1{#1}   \fi
\ifx \showURL      \undefined \def \showURL       {\relax}        \fi
% invisible to TeX
\providecommand\bibfield[2]{#2}
\providecommand\bibinfo[2]{#2}
\providecommand\natexlab[1]{#1}
\providecommand\showeprint[2][]{arXiv:#2}

\bibitem[\protect\citeauthoryear{Baltrunas, Ludwig, and Ricci}{Baltrunas
  et~al\mbox{.}}{2011}]%
        {camf}
\bibfield{author}{\bibinfo{person}{Linas Baltrunas}, \bibinfo{person}{Bernd
  Ludwig}, {and} \bibinfo{person}{Francesco Ricci}.}
  \bibinfo{year}{2011}\natexlab{}.
\newblock \showarticletitle{Matrix factorization techniques for context aware
  recommendation}. In \bibinfo{booktitle}{\emph{Proceedings of the fifth ACM
  conference on Recommender systems}}. ACM, \bibinfo{pages}{301--304}.
\newblock


\bibitem[\protect\citeauthoryear{Beutel, Covington, Jain, Xu, Li, Gatto, and
  Chi}{Beutel et~al\mbox{.}}{2018}]%
        {mult-google}
\bibfield{author}{\bibinfo{person}{Alex Beutel}, \bibinfo{person}{Paul
  Covington}, \bibinfo{person}{Sagar Jain}, \bibinfo{person}{Can Xu},
  \bibinfo{person}{Jia Li}, \bibinfo{person}{Vince Gatto}, {and}
  \bibinfo{person}{Ed~H Chi}.} \bibinfo{year}{2018}\natexlab{}.
\newblock \showarticletitle{Latent cross: Making use of context in recurrent
  recommender systems}. In \bibinfo{booktitle}{\emph{Proceedings of the
  Eleventh ACM International Conference on Web Search and Data Mining}}. ACM,
  \bibinfo{pages}{46--54}.
\newblock


\bibitem[\protect\citeauthoryear{Chang, Learned-Miller, and McCallum}{Chang
  et~al\mbox{.}}{2017}]%
        {active_bias}
\bibfield{author}{\bibinfo{person}{Haw-Shiuan Chang}, \bibinfo{person}{Erik
  Learned-Miller}, {and} \bibinfo{person}{Andrew McCallum}.}
  \bibinfo{year}{2017}\natexlab{}.
\newblock \showarticletitle{Active bias: Training more accurate neural networks
  by emphasizing high variance samples}. In \bibinfo{booktitle}{\emph{Advances
  in Neural Information Processing Systems}}. \bibinfo{pages}{1002--1012}.
\newblock


\bibitem[\protect\citeauthoryear{Chen, Dong, Li, and He}{Chen
  et~al\mbox{.}}{2018}]%
        {federated-meta}
\bibfield{author}{\bibinfo{person}{Fei Chen}, \bibinfo{person}{Zhenhua Dong},
  \bibinfo{person}{Zhenguo Li}, {and} \bibinfo{person}{Xiuqiang He}.}
  \bibinfo{year}{2018}\natexlab{}.
\newblock \showarticletitle{Federated meta-learning for recommendation}.
\newblock \bibinfo{journal}{\emph{arXiv preprint arXiv:1802.07876}}
  (\bibinfo{year}{2018}).
\newblock


\bibitem[\protect\citeauthoryear{Cogswell, Ahmed, Girshick, Zitnick, and
  Batra}{Cogswell et~al\mbox{.}}{2015}]%
        {decorrelate}
\bibfield{author}{\bibinfo{person}{Michael Cogswell}, \bibinfo{person}{Faruk
  Ahmed}, \bibinfo{person}{Ross Girshick}, \bibinfo{person}{Larry Zitnick},
  {and} \bibinfo{person}{Dhruv Batra}.} \bibinfo{year}{2015}\natexlab{}.
\newblock \showarticletitle{Reducing overfitting in deep networks by
  decorrelating representations}.
\newblock \bibinfo{journal}{\emph{arXiv preprint arXiv:1511.06068}}
  (\bibinfo{year}{2015}).
\newblock


\bibitem[\protect\citeauthoryear{Doersch}{Doersch}{2016}]%
        {vae-tutorial}
\bibfield{author}{\bibinfo{person}{Carl Doersch}.}
  \bibinfo{year}{2016}\natexlab{}.
\newblock \showarticletitle{Tutorial on variational autoencoders}.
\newblock \bibinfo{journal}{\emph{arXiv preprint arXiv:1606.05908}}
  (\bibinfo{year}{2016}).
\newblock


\bibitem[\protect\citeauthoryear{Du, Wang, Yang, Zhou, and Tang}{Du
  et~al\mbox{.}}{2019}]%
        {kdd19}
\bibfield{author}{\bibinfo{person}{Zhengxiao Du}, \bibinfo{person}{Xiaowei
  Wang}, \bibinfo{person}{Hongxia Yang}, \bibinfo{person}{Jingren Zhou}, {and}
  \bibinfo{person}{Jie Tang}.} \bibinfo{year}{2019}\natexlab{}.
\newblock \showarticletitle{Sequential Scenario-Specific Meta Learner for
  Online Recommendation}.
\newblock \bibinfo{journal}{\emph{arXiv preprint arXiv:1906.00391}}
  (\bibinfo{year}{2019}).
\newblock


\bibitem[\protect\citeauthoryear{Feurer, Springenberg, and Hutter}{Feurer
  et~al\mbox{.}}{2015}]%
        {metaparam}
\bibfield{author}{\bibinfo{person}{Matthias Feurer},
  \bibinfo{person}{Jost~Tobias Springenberg}, {and} \bibinfo{person}{Frank
  Hutter}.} \bibinfo{year}{2015}\natexlab{}.
\newblock \showarticletitle{Initializing bayesian hyperparameter optimization
  via meta-learning}. In \bibinfo{booktitle}{\emph{Twenty-Ninth AAAI Conference
  on Artificial Intelligence}}.
\newblock


\bibitem[\protect\citeauthoryear{Finn, Abbeel, and Levine}{Finn
  et~al\mbox{.}}{2017}]%
        {maml}
\bibfield{author}{\bibinfo{person}{Chelsea Finn}, \bibinfo{person}{Pieter
  Abbeel}, {and} \bibinfo{person}{Sergey Levine}.}
  \bibinfo{year}{2017}\natexlab{}.
\newblock \showarticletitle{Model-agnostic meta-learning for fast adaptation of
  deep networks}. In \bibinfo{booktitle}{\emph{Proceedings of the 34th
  International Conference on Machine Learning-Volume 70}}. JMLR. org,
  \bibinfo{pages}{1126--1135}.
\newblock


\bibitem[\protect\citeauthoryear{Gao, Chen, Feng, Zhao, He, Li, and Jin}{Gao
  et~al\mbox{.}}{2019}]%
        {ncf-privacy}
\bibfield{author}{\bibinfo{person}{Chen Gao}, \bibinfo{person}{Xiangning Chen},
  \bibinfo{person}{Fuli Feng}, \bibinfo{person}{Kai Zhao},
  \bibinfo{person}{Xiangnan He}, \bibinfo{person}{Yong Li}, {and}
  \bibinfo{person}{Depeng Jin}.} \bibinfo{year}{2019}\natexlab{}.
\newblock \showarticletitle{Cross-domain Recommendation Without Sharing
  User-relevant Data}. In \bibinfo{booktitle}{\emph{The World Wide Web
  Conference}}. ACM, \bibinfo{pages}{491--502}.
\newblock


\bibitem[\protect\citeauthoryear{Gao, Luo, Chen, Li, Gallinari, and Guo}{Gao
  et~al\mbox{.}}{2013}]%
        {ecmlpkdd13}
\bibfield{author}{\bibinfo{person}{Sheng Gao}, \bibinfo{person}{Hao Luo},
  \bibinfo{person}{Da Chen}, \bibinfo{person}{Shantao Li},
  \bibinfo{person}{Patrick Gallinari}, {and} \bibinfo{person}{Jun Guo}.}
  \bibinfo{year}{2013}\natexlab{}.
\newblock \showarticletitle{Cross-domain recommendation via cluster-level
  latent factor model}. In \bibinfo{booktitle}{\emph{Joint European conference
  on machine learning and knowledge discovery in databases}}. Springer,
  \bibinfo{pages}{161--176}.
\newblock


\bibitem[\protect\citeauthoryear{He, Zhang, Ren, and Sun}{He
  et~al\mbox{.}}{2016}]%
        {resnet}
\bibfield{author}{\bibinfo{person}{Kaiming He}, \bibinfo{person}{Xiangyu
  Zhang}, \bibinfo{person}{Shaoqing Ren}, {and} \bibinfo{person}{Jian Sun}.}
  \bibinfo{year}{2016}\natexlab{}.
\newblock \showarticletitle{Deep residual learning for image recognition}. In
  \bibinfo{booktitle}{\emph{Proceedings of the IEEE conference on computer
  vision and pattern recognition}}. \bibinfo{pages}{770--778}.
\newblock


\bibitem[\protect\citeauthoryear{He, Kang, and McAuley}{He
  et~al\mbox{.}}{2017a}]%
        {mcauley1}
\bibfield{author}{\bibinfo{person}{Ruining He}, \bibinfo{person}{Wang-Cheng
  Kang}, {and} \bibinfo{person}{Julian McAuley}.}
  \bibinfo{year}{2017}\natexlab{a}.
\newblock \showarticletitle{Translation-based recommendation}. In
  \bibinfo{booktitle}{\emph{Proceedings of the Eleventh ACM Conference on
  Recommender Systems}}. ACM, \bibinfo{pages}{161--169}.
\newblock


\bibitem[\protect\citeauthoryear{He and Chua}{He and Chua}{2017}]%
        {nfm}
\bibfield{author}{\bibinfo{person}{Xiangnan He} {and} \bibinfo{person}{Tat-Seng
  Chua}.} \bibinfo{year}{2017}\natexlab{}.
\newblock \showarticletitle{Neural factorization machines for sparse predictive
  analytics}. In \bibinfo{booktitle}{\emph{Proceedings of the 40th
  International ACM SIGIR conference on Research and Development in Information
  Retrieval}}. ACM, \bibinfo{pages}{355--364}.
\newblock


\bibitem[\protect\citeauthoryear{He, Liao, Zhang, Nie, Hu, and Chua}{He
  et~al\mbox{.}}{2017b}]%
        {ncf}
\bibfield{author}{\bibinfo{person}{Xiangnan He}, \bibinfo{person}{Lizi Liao},
  \bibinfo{person}{Hanwang Zhang}, \bibinfo{person}{Liqiang Nie},
  \bibinfo{person}{Xia Hu}, {and} \bibinfo{person}{Tat-Seng Chua}.}
  \bibinfo{year}{2017}\natexlab{b}.
\newblock \showarticletitle{Neural collaborative filtering}. In
  \bibinfo{booktitle}{\emph{Proceedings of the 26th international conference on
  world wide web}}. International World Wide Web Conferences Steering
  Committee, \bibinfo{pages}{173--182}.
\newblock


\bibitem[\protect\citeauthoryear{Hu, Zhang, and Yang}{Hu et~al\mbox{.}}{2019}]%
        {www19tmh}
\bibfield{author}{\bibinfo{person}{Guangneng Hu}, \bibinfo{person}{Yu Zhang},
  {and} \bibinfo{person}{Qiang Yang}.} \bibinfo{year}{2019}\natexlab{}.
\newblock \showarticletitle{Transfer Meets Hybrid: A Synthetic Approach for
  Cross-Domain Collaborative Filtering with Text}. In
  \bibinfo{booktitle}{\emph{The World Wide Web Conference}}. ACM,
  \bibinfo{pages}{2822--2829}.
\newblock


\bibitem[\protect\citeauthoryear{Jiang, Cui, Wang, Xu, Zhu, and Yang}{Jiang
  et~al\mbox{.}}{2014}]%
        {fema}
\bibfield{author}{\bibinfo{person}{Meng Jiang}, \bibinfo{person}{Peng Cui},
  \bibinfo{person}{Fei Wang}, \bibinfo{person}{Xinran Xu},
  \bibinfo{person}{Wenwu Zhu}, {and} \bibinfo{person}{Shiqiang Yang}.}
  \bibinfo{year}{2014}\natexlab{}.
\newblock \showarticletitle{Fema: flexible evolutionary multi-faceted analysis
  for dynamic behavioral pattern discovery}. In
  \bibinfo{booktitle}{\emph{Proceedings of the 20th ACM SIGKDD international
  conference on Knowledge discovery and data mining}}. ACM,
  \bibinfo{pages}{1186--1195}.
\newblock


\bibitem[\protect\citeauthoryear{Jiang, Ding, and Fu}{Jiang
  et~al\mbox{.}}{2017}]%
        {joint-factor1}
\bibfield{author}{\bibinfo{person}{Shuhui Jiang}, \bibinfo{person}{Zhengming
  Ding}, {and} \bibinfo{person}{Yun Fu}.} \bibinfo{year}{2017}\natexlab{}.
\newblock \showarticletitle{Deep low-rank sparse collective factorization for
  cross-domain recommendation}. In \bibinfo{booktitle}{\emph{Proceedings of the
  25th ACM international conference on Multimedia}}. ACM,
  \bibinfo{pages}{163--171}.
\newblock


\bibitem[\protect\citeauthoryear{Johnson and Zhang}{Johnson and Zhang}{2013}]%
        {var_reduction}
\bibfield{author}{\bibinfo{person}{Rie Johnson} {and} \bibinfo{person}{Tong
  Zhang}.} \bibinfo{year}{2013}\natexlab{}.
\newblock \showarticletitle{Accelerating stochastic gradient descent using
  predictive variance reduction}. In \bibinfo{booktitle}{\emph{Advances in
  neural information processing systems}}. \bibinfo{pages}{315--323}.
\newblock


\bibitem[\protect\citeauthoryear{Karatzoglou, Amatriain, Baltrunas, and
  Oliver}{Karatzoglou et~al\mbox{.}}{2010}]%
        {ctf}
\bibfield{author}{\bibinfo{person}{Alexandros Karatzoglou},
  \bibinfo{person}{Xavier Amatriain}, \bibinfo{person}{Linas Baltrunas}, {and}
  \bibinfo{person}{Nuria Oliver}.} \bibinfo{year}{2010}\natexlab{}.
\newblock \showarticletitle{Multiverse recommendation: n-dimensional tensor
  factorization for context-aware collaborative filtering}. In
  \bibinfo{booktitle}{\emph{Proceedings of the fourth ACM conference on
  Recommender systems}}. ACM, \bibinfo{pages}{79--86}.
\newblock


\bibitem[\protect\citeauthoryear{Kim, On, Lim, Kim, Ha, and Zhang}{Kim
  et~al\mbox{.}}{2016}]%
        {bilinear}
\bibfield{author}{\bibinfo{person}{Jin-Hwa Kim}, \bibinfo{person}{Kyoung-Woon
  On}, \bibinfo{person}{Woosang Lim}, \bibinfo{person}{Jeonghee Kim},
  \bibinfo{person}{Jung-Woo Ha}, {and} \bibinfo{person}{Byoung-Tak Zhang}.}
  \bibinfo{year}{2016}\natexlab{}.
\newblock \showarticletitle{Hadamard product for low-rank bilinear pooling}.
\newblock \bibinfo{journal}{\emph{arXiv preprint arXiv:1610.04325}}
  (\bibinfo{year}{2016}).
\newblock


\bibitem[\protect\citeauthoryear{Kingma and Ba}{Kingma and Ba}{2014}]%
        {adam}
\bibfield{author}{\bibinfo{person}{Diederik~P Kingma} {and}
  \bibinfo{person}{Jimmy Ba}.} \bibinfo{year}{2014}\natexlab{}.
\newblock \showarticletitle{Adam: A method for stochastic optimization}.
\newblock \bibinfo{journal}{\emph{arXiv preprint arXiv:1412.6980}}
  (\bibinfo{year}{2014}).
\newblock


\bibitem[\protect\citeauthoryear{Kirkpatrick, Gelatt, and Vecchi}{Kirkpatrick
  et~al\mbox{.}}{1983}]%
        {anneal}
\bibfield{author}{\bibinfo{person}{Scott Kirkpatrick},
  \bibinfo{person}{C~Daniel Gelatt}, {and} \bibinfo{person}{Mario~P Vecchi}.}
  \bibinfo{year}{1983}\natexlab{}.
\newblock \showarticletitle{Optimization by simulated annealing}.
\newblock \bibinfo{journal}{\emph{science}} \bibinfo{volume}{220},
  \bibinfo{number}{4598} (\bibinfo{year}{1983}), \bibinfo{pages}{671--680}.
\newblock


\bibitem[\protect\citeauthoryear{Krishnan, Cheruvu, Tao, and Sundaram}{Krishnan
  et~al\mbox{.}}{2019}]%
        {socialrec}
\bibfield{author}{\bibinfo{person}{Adit Krishnan}, \bibinfo{person}{Hari
  Cheruvu}, \bibinfo{person}{Cheng Tao}, {and} \bibinfo{person}{Hari
  Sundaram}.} \bibinfo{year}{2019}\natexlab{}.
\newblock \showarticletitle{A modular adversarial approach to social
  recommendation}. In \bibinfo{booktitle}{\emph{Proceedings of the 28th ACM
  International Conference on Information and Knowledge Management}}.
  \bibinfo{pages}{1753--1762}.
\newblock


\bibitem[\protect\citeauthoryear{Krishnan, Sharma, Sankar, and
  Sundaram}{Krishnan et~al\mbox{.}}{2018}]%
        {longtail-neural}
\bibfield{author}{\bibinfo{person}{Adit Krishnan}, \bibinfo{person}{Ashish
  Sharma}, \bibinfo{person}{Aravind Sankar}, {and} \bibinfo{person}{Hari
  Sundaram}.} \bibinfo{year}{2018}\natexlab{}.
\newblock \showarticletitle{An Adversarial Approach to Improve Long-Tail
  Performance in Neural Collaborative Filtering}. In
  \bibinfo{booktitle}{\emph{Proceedings of the 27th ACM International
  Conference on Information and Knowledge Management}}. ACM,
  \bibinfo{pages}{1491--1494}.
\newblock


\bibitem[\protect\citeauthoryear{Lee, Im, Jang, Cho, and Chung}{Lee
  et~al\mbox{.}}{2019}]%
        {melu}
\bibfield{author}{\bibinfo{person}{Hoyeop Lee}, \bibinfo{person}{Jinbae Im},
  \bibinfo{person}{Seongwon Jang}, \bibinfo{person}{Hyunsouk Cho}, {and}
  \bibinfo{person}{Sehee Chung}.} \bibinfo{year}{2019}\natexlab{}.
\newblock \showarticletitle{MeLU: Meta-Learned User Preference Estimator for
  Cold-Start Recommendation}. In \bibinfo{booktitle}{\emph{Proceedings of the
  25th ACM SIGKDD International Conference on Knowledge Discovery \& Data
  Mining}}. \bibinfo{pages}{1073--1082}.
\newblock


\bibitem[\protect\citeauthoryear{Li, Yang, and Xue}{Li et~al\mbox{.}}{2009}]%
        {codebook}
\bibfield{author}{\bibinfo{person}{Bin Li}, \bibinfo{person}{Qiang Yang}, {and}
  \bibinfo{person}{Xiangyang Xue}.} \bibinfo{year}{2009}\natexlab{}.
\newblock \showarticletitle{Can movies and books collaborate? cross-domain
  collaborative filtering for sparsity reduction}. In
  \bibinfo{booktitle}{\emph{Twenty-First International Joint Conference on
  Artificial Intelligence}}.
\newblock


\bibitem[\protect\citeauthoryear{Li, Zhou, Chen, and Li}{Li
  et~al\mbox{.}}{2017}]%
        {meta-learn-similar}
\bibfield{author}{\bibinfo{person}{Zhenguo Li}, \bibinfo{person}{Fengwei Zhou},
  \bibinfo{person}{Fei Chen}, {and} \bibinfo{person}{Hang Li}.}
  \bibinfo{year}{2017}\natexlab{}.
\newblock \showarticletitle{Meta-SGD: Learning to learn quickly for few-shot
  learning}.
\newblock \bibinfo{journal}{\emph{arXiv preprint arXiv:1707.09835}}
  (\bibinfo{year}{2017}).
\newblock


\bibitem[\protect\citeauthoryear{Liang, Krishnan, Hoffman, and Jebara}{Liang
  et~al\mbox{.}}{2018}]%
        {vaecf}
\bibfield{author}{\bibinfo{person}{Dawen Liang}, \bibinfo{person}{Rahul~G
  Krishnan}, \bibinfo{person}{Matthew~D Hoffman}, {and} \bibinfo{person}{Tony
  Jebara}.} \bibinfo{year}{2018}\natexlab{}.
\newblock \showarticletitle{Variational autoencoders for collaborative
  filtering}. In \bibinfo{booktitle}{\emph{Proceedings of the 2018 World Wide
  Web Conference}}. International World Wide Web Conferences Steering
  Committee, \bibinfo{pages}{689--698}.
\newblock


\bibitem[\protect\citeauthoryear{Liu, Wang, Gao, and Han}{Liu
  et~al\mbox{.}}{2013}]%
        {joint-factor2}
\bibfield{author}{\bibinfo{person}{Jialu Liu}, \bibinfo{person}{Chi Wang},
  \bibinfo{person}{Jing Gao}, {and} \bibinfo{person}{Jiawei Han}.}
  \bibinfo{year}{2013}\natexlab{}.
\newblock \showarticletitle{Multi-view clustering via joint nonnegative matrix
  factorization}. In \bibinfo{booktitle}{\emph{Proceedings of the 2013 SIAM
  International Conference on Data Mining}}. SIAM, \bibinfo{pages}{252--260}.
\newblock


\bibitem[\protect\citeauthoryear{Long, Zhu, Wang, and Jordan}{Long
  et~al\mbox{.}}{2016}]%
        {classifier-adap}
\bibfield{author}{\bibinfo{person}{Mingsheng Long}, \bibinfo{person}{Han Zhu},
  \bibinfo{person}{Jianmin Wang}, {and} \bibinfo{person}{Michael~I Jordan}.}
  \bibinfo{year}{2016}\natexlab{}.
\newblock \showarticletitle{Unsupervised domain adaptation with residual
  transfer networks}. In \bibinfo{booktitle}{\emph{Advances in Neural
  Information Processing Systems}}. \bibinfo{pages}{136--144}.
\newblock


\bibitem[\protect\citeauthoryear{Long, Zhu, Wang, and Jordan}{Long
  et~al\mbox{.}}{2017}]%
        {joint-adaptation-nets}
\bibfield{author}{\bibinfo{person}{Mingsheng Long}, \bibinfo{person}{Han Zhu},
  \bibinfo{person}{Jianmin Wang}, {and} \bibinfo{person}{Michael~I Jordan}.}
  \bibinfo{year}{2017}\natexlab{}.
\newblock \showarticletitle{Deep transfer learning with joint adaptation
  networks}. In \bibinfo{booktitle}{\emph{Proceedings of the 34th International
  Conference on Machine Learning-Volume 70}}. JMLR. org,
  \bibinfo{pages}{2208--2217}.
\newblock


\bibitem[\protect\citeauthoryear{Loshchilov and Hutter}{Loshchilov and
  Hutter}{2015}]%
        {hardsample1}
\bibfield{author}{\bibinfo{person}{Ilya Loshchilov} {and}
  \bibinfo{person}{Frank Hutter}.} \bibinfo{year}{2015}\natexlab{}.
\newblock \showarticletitle{Online batch selection for faster training of
  neural networks}.
\newblock \bibinfo{journal}{\emph{arXiv preprint arXiv:1511.06343}}
  (\bibinfo{year}{2015}).
\newblock


\bibitem[\protect\citeauthoryear{Man, Shen, Jin, and Cheng}{Man
  et~al\mbox{.}}{2017}]%
        {ijcai17}
\bibfield{author}{\bibinfo{person}{Tong Man}, \bibinfo{person}{Huawei Shen},
  \bibinfo{person}{Xiaolong Jin}, {and} \bibinfo{person}{Xueqi Cheng}.}
  \bibinfo{year}{2017}\natexlab{}.
\newblock \showarticletitle{Cross-Domain Recommendation: An Embedding and
  Mapping Approach.}. In \bibinfo{booktitle}{\emph{IJCAI}}.
  \bibinfo{pages}{2464--2470}.
\newblock


\bibitem[\protect\citeauthoryear{Mei, Ren, Chen, Nie, Ma, and Nie}{Mei
  et~al\mbox{.}}{2018}]%
        {cikm18context}
\bibfield{author}{\bibinfo{person}{Lei Mei}, \bibinfo{person}{Pengjie Ren},
  \bibinfo{person}{Zhumin Chen}, \bibinfo{person}{Liqiang Nie},
  \bibinfo{person}{Jun Ma}, {and} \bibinfo{person}{Jian-Yun Nie}.}
  \bibinfo{year}{2018}\natexlab{}.
\newblock \showarticletitle{An attentive interaction network for context-aware
  recommendations}. In \bibinfo{booktitle}{\emph{Proceedings of the 27th ACM
  International Conference on Information and Knowledge Management}}. ACM,
  \bibinfo{pages}{157--166}.
\newblock


\bibitem[\protect\citeauthoryear{Pan, Xiang, Liu, and Yang}{Pan
  et~al\mbox{.}}{2010}]%
        {aaai10}
\bibfield{author}{\bibinfo{person}{Weike Pan}, \bibinfo{person}{Evan~Wei
  Xiang}, \bibinfo{person}{Nathan~Nan Liu}, {and} \bibinfo{person}{Qiang
  Yang}.} \bibinfo{year}{2010}\natexlab{}.
\newblock \showarticletitle{Transfer learning in collaborative filtering for
  sparsity reduction}. In \bibinfo{booktitle}{\emph{Twenty-fourth AAAI
  conference on artificial intelligence}}.
\newblock


\bibitem[\protect\citeauthoryear{Pasricha and McAuley}{Pasricha and
  McAuley}{2018}]%
        {mcauley2}
\bibfield{author}{\bibinfo{person}{Rajiv Pasricha} {and}
  \bibinfo{person}{Julian McAuley}.} \bibinfo{year}{2018}\natexlab{}.
\newblock \showarticletitle{Translation-based factorization machines for
  sequential recommendation}. In \bibinfo{booktitle}{\emph{Proceedings of the
  12th ACM Conference on Recommender Systems}}. ACM, \bibinfo{pages}{63--71}.
\newblock


\bibitem[\protect\citeauthoryear{Perera and Zimmermann}{Perera and
  Zimmermann}{2019}]%
        {www19cngan}
\bibfield{author}{\bibinfo{person}{Dilruk Perera} {and} \bibinfo{person}{Roger
  Zimmermann}.} \bibinfo{year}{2019}\natexlab{}.
\newblock \showarticletitle{CnGAN: Generative Adversarial Networks for
  Cross-network user preference generation for non-overlapped users}. In
  \bibinfo{booktitle}{\emph{The World Wide Web Conference}}. ACM,
  \bibinfo{pages}{3144--3150}.
\newblock


\bibitem[\protect\citeauthoryear{Radford, Metz, and Chintala}{Radford
  et~al\mbox{.}}{2015}]%
        {adv-dist}
\bibfield{author}{\bibinfo{person}{Alec Radford}, \bibinfo{person}{Luke Metz},
  {and} \bibinfo{person}{Soumith Chintala}.} \bibinfo{year}{2015}\natexlab{}.
\newblock \showarticletitle{Unsupervised representation learning with deep
  convolutional generative adversarial networks}.
\newblock \bibinfo{journal}{\emph{arXiv preprint arXiv:1511.06434}}
  (\bibinfo{year}{2015}).
\newblock


\bibitem[\protect\citeauthoryear{Sabokrou, Khalooei, Fathy, and Adeli}{Sabokrou
  et~al\mbox{.}}{2018}]%
        {adv-novelty}
\bibfield{author}{\bibinfo{person}{Mohammad Sabokrou},
  \bibinfo{person}{Mohammad Khalooei}, \bibinfo{person}{Mahmood Fathy}, {and}
  \bibinfo{person}{Ehsan Adeli}.} \bibinfo{year}{2018}\natexlab{}.
\newblock \showarticletitle{Adversarially learned one-class classifier for
  novelty detection}. In \bibinfo{booktitle}{\emph{Proceedings of the IEEE
  Conference on Computer Vision and Pattern Recognition}}.
  \bibinfo{pages}{3379--3388}.
\newblock


\bibitem[\protect\citeauthoryear{Sankar, Zhang, and Chang}{Sankar
  et~al\mbox{.}}{2019}]%
        {motif}
\bibfield{author}{\bibinfo{person}{Aravind Sankar}, \bibinfo{person}{Xinyang
  Zhang}, {and} \bibinfo{person}{Kevin Chen-Chuan Chang}.}
  \bibinfo{year}{2019}\natexlab{}.
\newblock \showarticletitle{Meta-GNN: metagraph neural network for
  semi-supervised learning in attributed heterogeneous information networks}.
  In \bibinfo{booktitle}{\emph{Proceedings of the 2019 IEEE/ACM International
  Conference on Advances in Social Networks Analysis and Mining}}.
  \bibinfo{pages}{137--144}.
\newblock


\bibitem[\protect\citeauthoryear{Sankar, Zhang, Krishnan, and Han}{Sankar
  et~al\mbox{.}}{2020}]%
        {infvae}
\bibfield{author}{\bibinfo{person}{Aravind Sankar}, \bibinfo{person}{Xinyang
  Zhang}, \bibinfo{person}{Adit Krishnan}, {and} \bibinfo{person}{Jiawei Han}.}
  \bibinfo{year}{2020}\natexlab{}.
\newblock \showarticletitle{Inf-VAE: A Variational Autoencoder Framework to
  Integrate Homophily and Influence in Diffusion Prediction}.
\newblock \bibinfo{journal}{\emph{arXiv preprint arXiv:2001.00132}}
  (\bibinfo{year}{2020}).
\newblock


\bibitem[\protect\citeauthoryear{Srivastava, Hinton, Krizhevsky, Sutskever, and
  Salakhutdinov}{Srivastava et~al\mbox{.}}{2014}]%
        {dropout}
\bibfield{author}{\bibinfo{person}{Nitish Srivastava},
  \bibinfo{person}{Geoffrey Hinton}, \bibinfo{person}{Alex Krizhevsky},
  \bibinfo{person}{Ilya Sutskever}, {and} \bibinfo{person}{Ruslan
  Salakhutdinov}.} \bibinfo{year}{2014}\natexlab{}.
\newblock \showarticletitle{Dropout: a simple way to prevent neural networks
  from overfitting}.
\newblock \bibinfo{journal}{\emph{The journal of machine learning research}}
  \bibinfo{volume}{15}, \bibinfo{number}{1} (\bibinfo{year}{2014}),
  \bibinfo{pages}{1929--1958}.
\newblock


\bibitem[\protect\citeauthoryear{Sun and Saenko}{Sun and Saenko}{2016}]%
        {layer-transfer-coral}
\bibfield{author}{\bibinfo{person}{Baochen Sun} {and} \bibinfo{person}{Kate
  Saenko}.} \bibinfo{year}{2016}\natexlab{}.
\newblock \showarticletitle{Deep coral: Correlation alignment for deep domain
  adaptation}. In \bibinfo{booktitle}{\emph{European Conference on Computer
  Vision}}. Springer, \bibinfo{pages}{443--450}.
\newblock


\bibitem[\protect\citeauthoryear{Sun, Liu, Chua, and Schiele}{Sun
  et~al\mbox{.}}{2018}]%
        {DBLP:journals/corr/abs-1812-02391}
\bibfield{author}{\bibinfo{person}{Qianru Sun}, \bibinfo{person}{Yaoyao Liu},
  \bibinfo{person}{Tat{-}Seng Chua}, {and} \bibinfo{person}{Bernt Schiele}.}
  \bibinfo{year}{2018}\natexlab{}.
\newblock \showarticletitle{Meta-Transfer Learning for Few-Shot Learning}.
\newblock \bibinfo{journal}{\emph{CoRR}}  \bibinfo{volume}{abs/1812.02391}
  (\bibinfo{year}{2018}).
\newblock
\showeprint[arxiv]{1812.02391}
\urldef\tempurl%
\url{http://arxiv.org/abs/1812.02391}
\showURL{%
\tempurl}


\bibitem[\protect\citeauthoryear{Tang, Alelyani, and Liu}{Tang
  et~al\mbox{.}}{2014}]%
        {featurepool}
\bibfield{author}{\bibinfo{person}{Jiliang Tang}, \bibinfo{person}{Salem
  Alelyani}, {and} \bibinfo{person}{Huan Liu}.}
  \bibinfo{year}{2014}\natexlab{}.
\newblock \showarticletitle{Feature selection for classification: A review}.
\newblock \bibinfo{journal}{\emph{Data classification: Algorithms and
  applications}} (\bibinfo{year}{2014}), \bibinfo{pages}{37}.
\newblock


\bibitem[\protect\citeauthoryear{Vartak, Thiagarajan, Miranda, Bratman, and
  Larochelle}{Vartak et~al\mbox{.}}{2017}]%
        {nips17}
\bibfield{author}{\bibinfo{person}{Manasi Vartak}, \bibinfo{person}{Arvind
  Thiagarajan}, \bibinfo{person}{Conrado Miranda}, \bibinfo{person}{Jeshua
  Bratman}, {and} \bibinfo{person}{Hugo Larochelle}.}
  \bibinfo{year}{2017}\natexlab{}.
\newblock \showarticletitle{A meta-learning perspective on cold-start
  recommendations for items}. In \bibinfo{booktitle}{\emph{Advances in neural
  information processing systems}}. \bibinfo{pages}{6904--6914}.
\newblock


\bibitem[\protect\citeauthoryear{Volkovs, Yu, and Poutanen}{Volkovs
  et~al\mbox{.}}{2017}]%
        {volkovs2017content}
\bibfield{author}{\bibinfo{person}{Maksims Volkovs}, \bibinfo{person}{Guang~Wei
  Yu}, {and} \bibinfo{person}{Tomi Poutanen}.} \bibinfo{year}{2017}\natexlab{}.
\newblock \showarticletitle{Content-based neighbor models for cold start in
  recommender systems}. In \bibinfo{booktitle}{\emph{Proceedings of the
  Recommender Systems Challenge 2017}}. ACM, \bibinfo{pages}{7}.
\newblock


\bibitem[\protect\citeauthoryear{Wang, Feng, Guo, Chu, and Hwang}{Wang
  et~al\mbox{.}}{2019}]%
        {wsdm19}
\bibfield{author}{\bibinfo{person}{Yaqing Wang}, \bibinfo{person}{Chunyan
  Feng}, \bibinfo{person}{Caili Guo}, \bibinfo{person}{Yunfei Chu}, {and}
  \bibinfo{person}{Jenq-Neng Hwang}.} \bibinfo{year}{2019}\natexlab{}.
\newblock \showarticletitle{Solving the Sparsity Problem in Recommendations via
  Cross-Domain Item Embedding Based on Co-Clustering}. In
  \bibinfo{booktitle}{\emph{Proceedings of the Twelfth ACM International
  Conference on Web Search and Data Mining}}. ACM, \bibinfo{pages}{717--725}.
\newblock


\bibitem[\protect\citeauthoryear{Xiao, Ye, He, Zhang, Wu, and Chua}{Xiao
  et~al\mbox{.}}{2017}]%
        {afm}
\bibfield{author}{\bibinfo{person}{Jun Xiao}, \bibinfo{person}{Hao Ye},
  \bibinfo{person}{Xiangnan He}, \bibinfo{person}{Hanwang Zhang},
  \bibinfo{person}{Fei Wu}, {and} \bibinfo{person}{Tat-Seng Chua}.}
  \bibinfo{year}{2017}\natexlab{}.
\newblock \showarticletitle{Attentional factorization machines: Learning the
  weight of feature interactions via attention networks}.
\newblock \bibinfo{journal}{\emph{arXiv preprint arXiv:1708.04617}}
  (\bibinfo{year}{2017}).
\newblock


\end{thebibliography}
